\documentclass[letterpaper,11pt]{article}

\usepackage[T1]{fontenc}
\usepackage[utf8]{inputenc}

\usepackage{graphicx}
\usepackage{epstopdf}
\usepackage{multicol}
\usepackage{lmodern}

\usepackage{jheppub}
\usepackage{microtype}   
\usepackage{multirow, amsthm}
\usepackage{braket}
\usepackage{mathrsfs}
\usepackage{tikz}
\usepackage[caption=false]{subfig}
\usepackage{xspace}
\usepackage{xcolor}
\usepackage{float,color}
\usepackage[separate-uncertainty=true]{siunitx}
\usepackage[ruled,norelsize]{algorithm2e}
\usepackage[countmax]{subfloat}
\usepackage{accents}
\usepackage{physics}
\usepackage[capitalise]{cleveref}    

\usepackage{tikz}
\usetikzlibrary{arrows,shapes}
\usetikzlibrary{trees}
\usetikzlibrary{matrix,arrows} 				
\usetikzlibrary{positioning}				
\usetikzlibrary{calc,through}				
\usetikzlibrary{decorations.pathreplacing}  
\usepackage{pgffor}							

\usetikzlibrary{decorations.pathmorphing}	
\usetikzlibrary{decorations.markings}
\usetikzlibrary{snakes}
\tikzset{
    vector/.style={decorate, decoration={snake}, draw},
	provector/.style={decorate, decoration={snake,amplitude=2.5pt}, draw},
	antivector/.style={decorate, decoration={snake,amplitude=-2.5pt}, draw},
    fermion/.style={draw=black, postaction={decorate},
        decoration={markings,mark=at position .55 with {\arrow[draw=black]{>}}}},
    fermionbar/.style={draw=black, postaction={decorate},
        decoration={markings,mark=at position .55 with {\arrow[draw=black]{<}}}},
    fermionnoarrow/.style={draw=black},
    gluon/.style={decorate, draw=black,
        decoration={coil,amplitude=4pt, segment length=5pt}},
    scalar/.style={dashed,draw=black, postaction={decorate},
        decoration={markings,mark=at position .55 with {\arrow[draw=black]{>}}}},
    scalarbar/.style={dashed,draw=black, postaction={decorate},
        decoration={markings,mark=at position .55 with {\arrow[draw=black]{<}}}},
    scalarnoarrow/.style={dashed,draw=black},
    electron/.style={draw=black, postaction={decorate},
        decoration={markings,mark=at position .55 with {\arrow[draw=black]{>}}}},
	bigvector/.style={decorate, decoration={snake,amplitude=4pt}, draw},
}

\tikzstyle{block} = [draw, rectangle, 
    minimum height=3em, minimum width=6em]

\newlength{\dhatheight}

\providecommand{\href}[2]{#2}
\usepackage{comment}

\definecolor{darkred}{rgb}{0.5,0.0,0.0}
\definecolor{darkblue}{rgb}{0.0,0.0,0.9}
\definecolor{darkerblue}{rgb}{0.0,0.0,0.5}
\definecolor{darkgreen}{rgb}{0.0,0.5,0.0}
\definecolor{black}{rgb}{0.0,0.0,0.0}
\definecolor{brown}{rgb}{0.6,0.4,0.2}

\DeclareSIUnit{\nb}{\nano\barn}
\DeclareSIUnit{\pb}{\pico\barn}
\DeclareSIUnit{\fb}{\femto\barn}
\DeclareSIUnit{\year}{yr}

\newcommand{\cathode}{\textsc{salad}}
\newcommand{\dctr}{\textsc{dctr}}

\title{\boldmath Simulation Assisted Likelihood-free Anomaly Detection
}
\author[1]{Anders Andreassen,}
\author[2]{Benjamin Nachman,}
\author[2,3,4]{and David Shih}
\affiliation[1]{\normalsize Google, Mountain View, CA 94043, USA}
\affiliation[2]{\normalsize Physics Division, Lawrence Berkeley National Laboratory, Berkeley, CA 94720, USA}
\affiliation[3]{\normalsize NHETC, Dept. of Physics and Astronomy, Rutgers, Piscataway, NJ 08854, USA}
\affiliation[4]{\normalsize Berkeley Center for Theoretical Physics, University of California, Berkeley, CA 94720, USA}

\emailAdd{ajandreassen@google.com}
\emailAdd{bpnachman@lbl.gov}
\emailAdd{shih@physics.rutgers.edu}

\abstract{
Given the lack of evidence for new particle discoveries at the Large Hadron Collider (LHC), it is critical to broaden the search program.  A variety of model-independent searches have been proposed, adding sensitivity to unexpected signals.   There are generally two types of such searches: those that rely heavily on simulations and those that are entirely based on (unlabeled) data.  This paper introduces a hybrid method that makes the best of both approaches.  For potential signals that are resonant in one known feature, this new method first learns a parameterized reweighting function to morph a given simulation to match the data in sidebands.  This function is then interpolated into the signal region and then the reweighted background-only simulation can be used for supervised learning as well as for background estimation.  The background estimation from the reweighted simulation allows for non-trivial correlations between features used for classification and the resonant feature.   A dijet search with jet substructure is used to illustrate the new method.  Future applications of Simulation Assisted Likelihood-free Anomaly Detection (\cathode) include a variety of final states and potential combinations with other model-independent approaches.
}

\begin{document} 
\maketitle
\flushbottom

\section{Introduction}
\label{sec:intro}

An immense search effort by the LHC collaborations has successfully probed many extreme regions of the Standard Model phase space~\cite{atlasexoticstwiki,atlassusytwiki,atlashdbspublictwiki,cmsexoticstwiki,cmssusytwiki,cmsb2gtwiki,lhcbtwiki}.  Despite strong theoretical and non-collider experimental motivation, there is currently no convincing evidence for new particles or forces of nature from the LHC searches.  However, many final states are uncovered~\cite{Kim:2019rhy,Craig:2016rqv} and the full hypervariate phase space accessible by modern detector technology is only starting to be probed holistically with deep learning methods~\cite{Larkoski:2017jix,Guest:2018yhq,Abdughani:2019wuv,Radovic:2018dip}.  There is a great need for new searches that can identify unexpected scenarios.


Until recently, nearly all model independent searches relied heavily on simulation.  Generically, these searches operate by comparing data with background-only simulation in a large number of phase space regions.  Such searches have been performed without machine learning at D0~\cite{sleuth,Abbott:2000fb,Abbott:2000gx,Abbott:2001ke}, H1~\cite{Aaron:2008aa,Aktas:2004pz}, CDF~\cite{Aaltonen:2007dg,Aaltonen:2007ab,Aaltonen:2008vt}, CMS~\cite{CMS-PAS-EXO-14-016,CMS-PAS-EXO-10-021}, and ATLAS~\cite{Aaboud:2018ufy,ATLAS-CONF-2014-006,ATLAS-CONF-2012-107}.  A recent phenomenological study proposed extending this idea to deep learning classifiers~\cite{DAgnolo:2018cun,DAgnolo:2019vbw}.  While independent of signal models, these approaches are dependent on the fidelity of the background model simulation for both signal sensitivity and background accuracy.  If the background simulation is inaccurate, then differences between simulation and (background-only) data will hide potential signals.  Even if a biased simulation can find a signal, if the background is mis-modeled, then the signal specificity will be poor. 

A variety of approaches have been proposed to enhance signal sensitivity without simulations.  Such proposals are based on clustering or nearest neighbor algorithms~\cite{DeSimone:2018efk,Mullin:2019mmh,1809.02977}, autoencoders~\cite{Farina:2018fyg,Heimel:2018mkt,Roy:2019jae,Cerri:2018anq,Blance:2019ibf,Hajer:2018kqm}, probabilistic modeling~\cite{Dillon:2019cqt}, weak supervision~\cite{Collins:2018epr,Collins:2019jip}, density estimation~\cite{anode}, and others~\cite{Aguilar-Saavedra:2017rzt}.  These approaches must also be combined with a background estimation strategy.  If simulation is used to estimate the background, then the specificity is the same as the model-dependent searches.  Many of these approaches can be combined with a resonance search, as explicitly demonstrated in Ref.~\cite{Collins:2018epr,Collins:2019jip}.  The background estimation strategy may impose additional constraints on the learning, such as the need for decorrelation between a resonant feature and other discriminative features~\cite{Louppe:2016ylz,Dolen:2016kst,Moult:2017okx,Stevens:2013dya,Shimmin:2017mfk,Bradshaw:2019ipy,ATL-PHYS-PUB-2018-014,Xia:2018kgd,Englert:2018cfo,Wunsch:2019qbo,Disco}.  A detailed overview of model independent approaches can be found in Ref.~\cite{anode}.

While it is desirable to be robust to background model inaccuracies, it is also useful to incorporate information from Standard Model simulations.  Even though these simulations are only an approximation to nature, they include an extensive set of fundamental and phenomenological physics models describing the highest energy reactions all the way to signal formation in the detector electronics.  This paper describes a method that uses a background simulation in a way that depends as little on that simulation as possible.  In particular, a model based on the \textit{Deep neural networks using Classification for Tuning and Reweighting} (\dctr) procedure~\cite{Andreassen:2019nnm} is trained in a region of phase space that is expected to be devoid of signals.  In a resonance search, there is one feature where the signal is known to be localized and the sideband can be used to train the \dctr~model.  This reweighting function learns to morph the simulation into the data and is parameterized in the feature(s) used to mask potential signals.  Then, the model is interpolated to the signal-sensitive region and the reweighted background simulation can be used for both enhancing signal sensitivity and estimating the Standard Model background.  As deep learning classifiers can naturally probe high dimensional spaces, this reweighting model can in principle exploit the full phase space for both enhancing signal sensitivity and specificity.

This paper is organized as follows.  Section~\ref{sec:methods} introduces the \textit{Simulation Assisted Likelihood-free Anomaly Detection} (\cathode) method.  A dijet search at the LHC is emulated to illustrate the new method.  The simulation and deep learning setup are introduced in Sec.~\ref{sec:sim} and then the application of \dctr~is shown in Sec.~\ref{sec:dctr}.   The signal sensitivity and specificity are presented in Sec.~\ref{sec:sens} and~\ref{sec:back}, respectively.   The paper ends with conclusions and outlook in Sec.~\ref{sec:conclusions}.

\clearpage

\section{Methods}
\label{sec:methods}

Let $m$ be a feature (or set of features) that can be used to localize a potential signal in a signal region (SR).  Furthermore, let $x$ be another set of features which are useful for isolating a potential signal.  The prototypical example is a resonance search where $m$ is the single resonant feature, such as the invariant mass of two jets, while $x$ are other properties of the event, such as the substructure of the two jets.   The \cathode~method then proceeds as follows:

\begin{enumerate}
\item Train a classifier $f$ to distinguish data and simulation for $m\not\in\text{SR}$.  This classifier is parameterized in $m$ by simply augmenting $x$ with $m$, $f=f(x,m)$~\cite{Cranmer:2015bka,Baldi:2016fzo}.  If $f$ is trained using the binary cross entropy or the mean squared error loss, then asymptotically, a weight function $w(x|m)$ is defined by 

\begin{align}
\label{eq:weightfunction}
w(x|m)\equiv \frac{f(x)}{1-f(x)} = \frac{p(x|\text{data})}{p(x|\text{simulation})}\times\frac{p(\text{data})}{p(\text{simulation})},
\end{align}

\noindent where the last factor in Eq.~\ref{eq:weightfunction} is an overall constant that is the ratio of the total amount of data to the total amount of simulation.  This property of neural networks to learn likelihood ratios has been exploited for a variety of full phase space reweighting and parameter estimation proposals in high energy physics~\cite{Andreassen:2019nnm,Brehmer:2018hga,Brehmer:2018eca,Brehmer:2018kdj,Cranmer:2015bka,Andreassen:2019cjw}.

\item Simulated events in the SR are reweighted using $w(x|m)$.  The function $w(x|m)$ is interpolated automatically by the neural network.  A second classifier $g(x)$ is used to distinguish the reweighted simulation from the data.  This can be achieved in the usual way with a weighted loss function such as the binary cross-entropy:

\begin{align}
\text{loss}(g(x))=-\sum_{m_i\in\text{SR}_\text{data}} \log g(x_i)-\sum_{m_i\in\text{SR}_\text{simulation}} w(x_i|m_i)\log (1-g(x_i)).
\end{align}

\noindent Events are then selected with large values of $g(x)$.  Asymptotically\footnote{Sufficiently flexible neural network architecture, enough training data, and an effective optimization procedure.}, $g(x)$ will be monotonically related with the optimal classifier:

\begin{align}
\frac{g(x)}{1-g(x)}\propto \frac{p(x|\text{signal+background})}{p(x|\text{background})}.
\end{align}

\noindent It is important that the same data are not used for training and testing.  The easiest way to achieve this is using different partitions of the data for these two tasks.  One can make use of more data with a cross-validation procedure~\cite{Collins:2018epr,Collins:2019jip}.

\item One could combine the previous step with a standard data-driven background estimation technique like a sideband fit or the ABCD method.  However, one can also directly use the weighted simulation to predict the number of events that should pass a threshold requirement on $g(x)$:

\begin{align}
\label{eq:backgroundprediction}
N_\text{predicted}(c)=\sum_{m_i\in\text{SR}_\text{simulation}} w(x_i|m_i)\mathbb{I}[g(x_i)>c],
\end{align}

for some threshold value $c$.  The advantage of Eq.~\ref{eq:backgroundprediction} over other data-based methods is that $g(x)$ could be correlated with $m$; for sideband fits, thresholds requirements on $g$ cannot sculpt local features in the $m$ spectrum.

\end{enumerate}

\section{Simulation}
\label{sec:sim}

A large-radius dijet resonance search is used to illustrate the \cathode~method.  The simulations are from the LHC Olympics 2020 community challenge R\&D dataset~\cite{gregor_kasieczka_2019_2629073}.  The background process is generic $2\rightarrow 2$ parton scattering (labeled QCD for quantum chromodynamics) and the signal is a hypothetical $W'$ boson that decays into an $X$ boson and  $Y$ boson.  Each of the $X$ and $Y$ decay to quarks.  The masses of the $W', X$, and $Y$ particles are 3.5, 0.5, and 0.1 TeV, respectively.  The mass hierarchy between the $W'$ particle and its decay products means that the $X$ and $Y$ particles are Lorentz boosted in the lab frame and therefore their two-prong decay products are captured inside a single large-radius jet.  Particle-level simulations are produced with Pythia~8~\cite{Sjostrand:2006za,Sjostrand:2007gs} or Herwig++~\cite{Bahr:2008pv} without pileup or multiple parton interactions and a detector simulation is performed with Delphes~3.4.1~\cite{deFavereau:2013fsa,Mertens:2015kba,Selvaggi:2014mya}. Particle flow objects are clustered into jets using the Fastjet~\cite{Cacciari:2011ma,Cacciari:2005hq} implementation of the anti-$k_t$ algorithm~\cite{Cacciari:2008gp} using $R=1.0$ as the jet radius.  Events are selected by requiring at least one such jet with $p_T>1.3$ TeV.  In the remaining studies, the Pythia dataset is treated as `data', while the Herwig dataset is treated as `simulation', to mimic the scenario in practice where the simulation is different than data.

\begin{figure}[h!]
\centering
\includegraphics[width=0.65\textwidth]{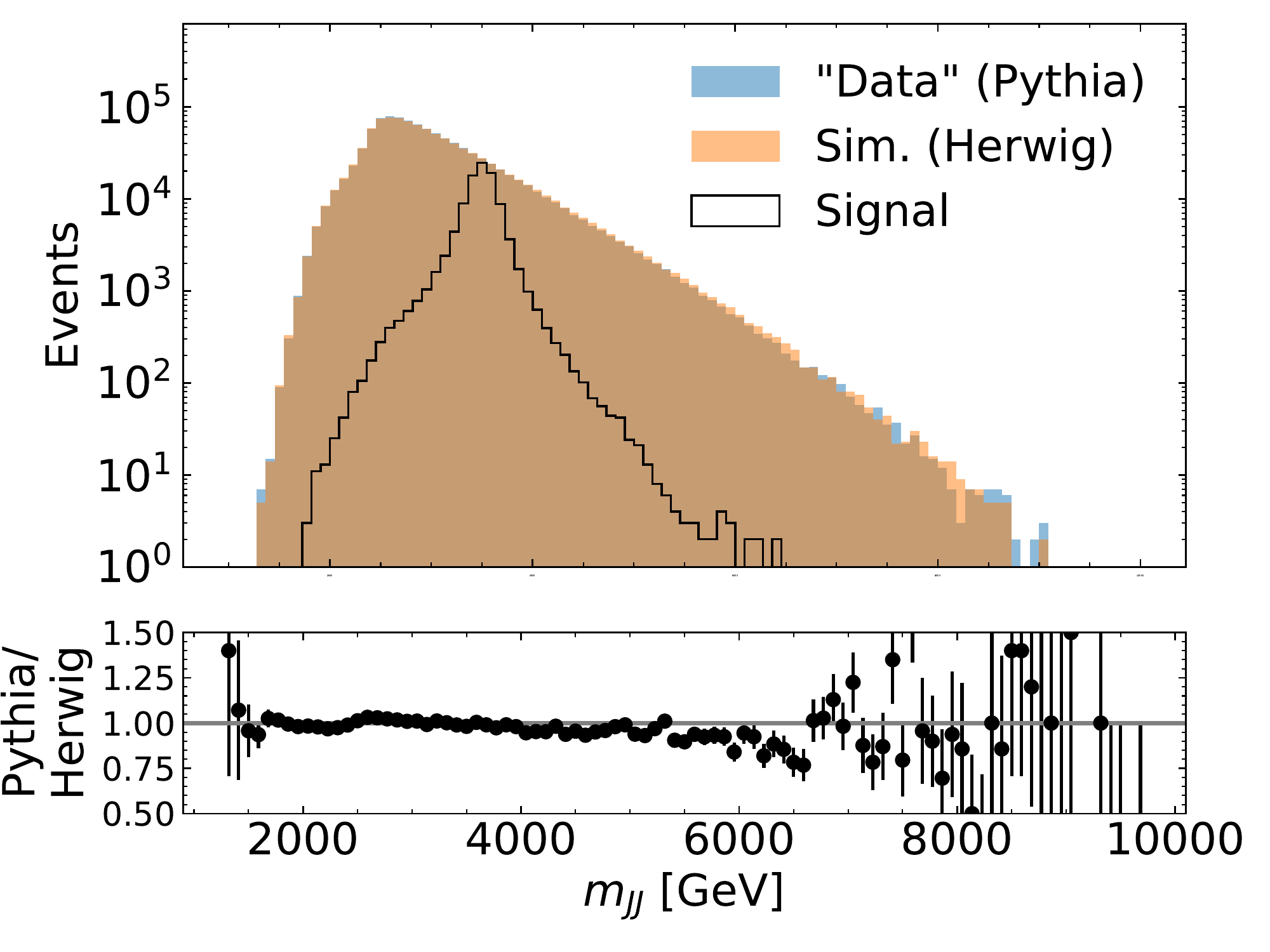}
\caption{The invariant mass of the leading two jets.}
\label{fig:mjj}
\end{figure}

Figure~\ref{fig:mjj} presents the invariant mass of the leading two jets.  The $p_T$ selection is evident from the peak around 3 TeV.  The signal peaks around the $W'$ mass and aside from the kinematic feature from the jet selection, the background distribution is featureless.  The spectra from Pythia and Herwig are nearly identical, which may be expected since the invariant mass is mostly determined by hard-scatter matrix elements and not final state effects.

\begin{figure}[h!]
\centering
\includegraphics[width=0.5\textwidth]{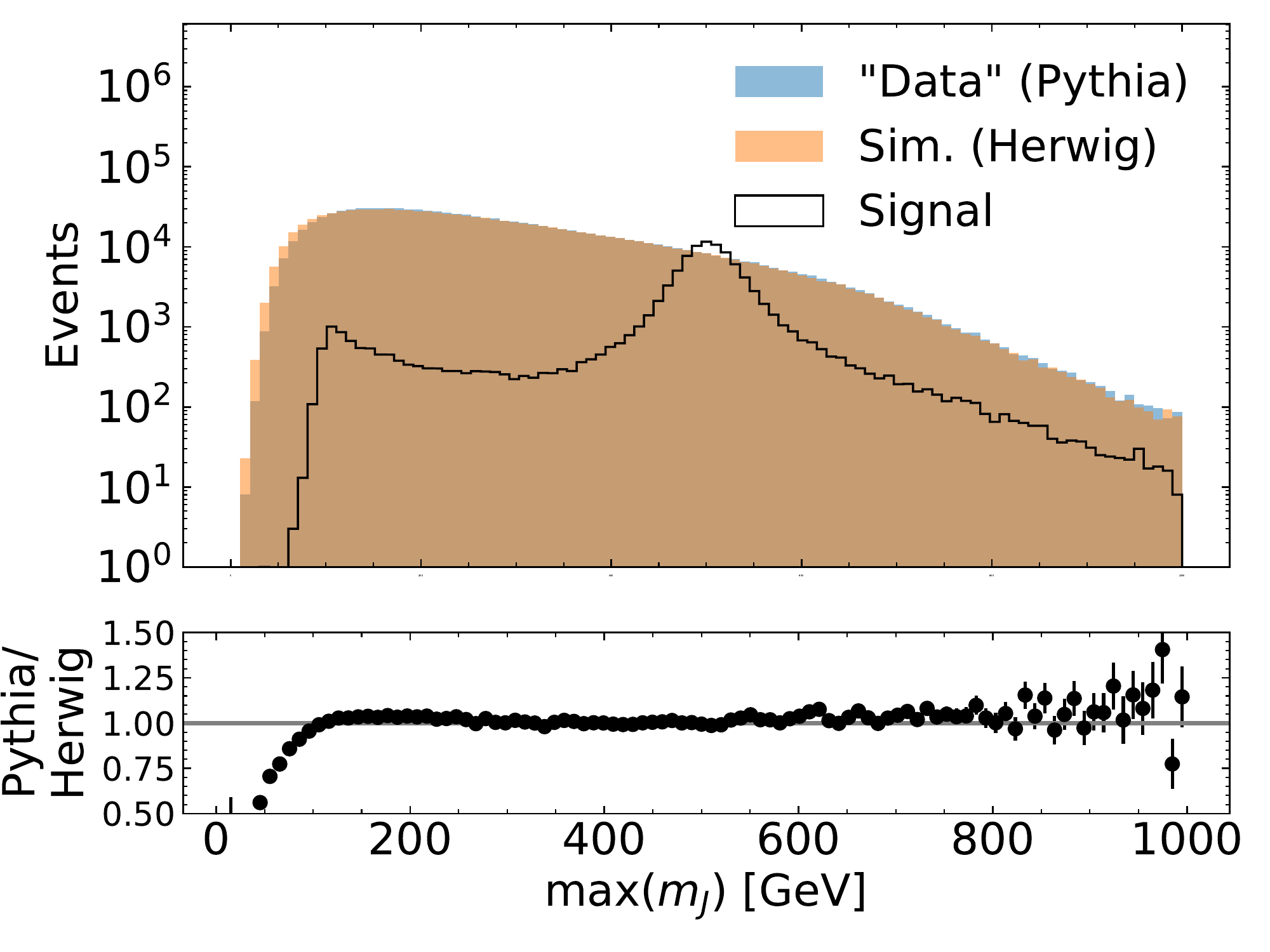}\includegraphics[width=0.5\textwidth]{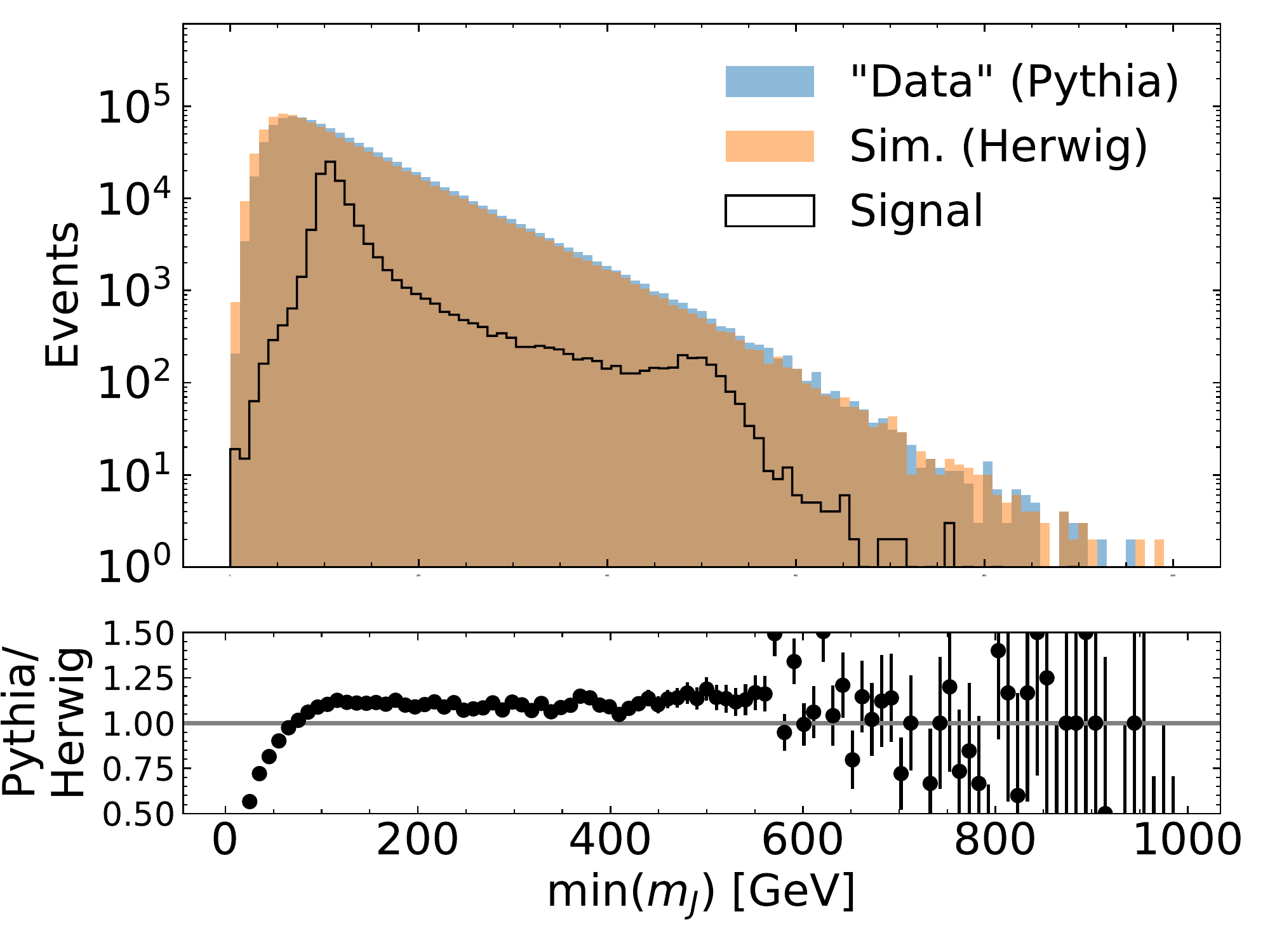}\\
\includegraphics[width=0.5\textwidth]{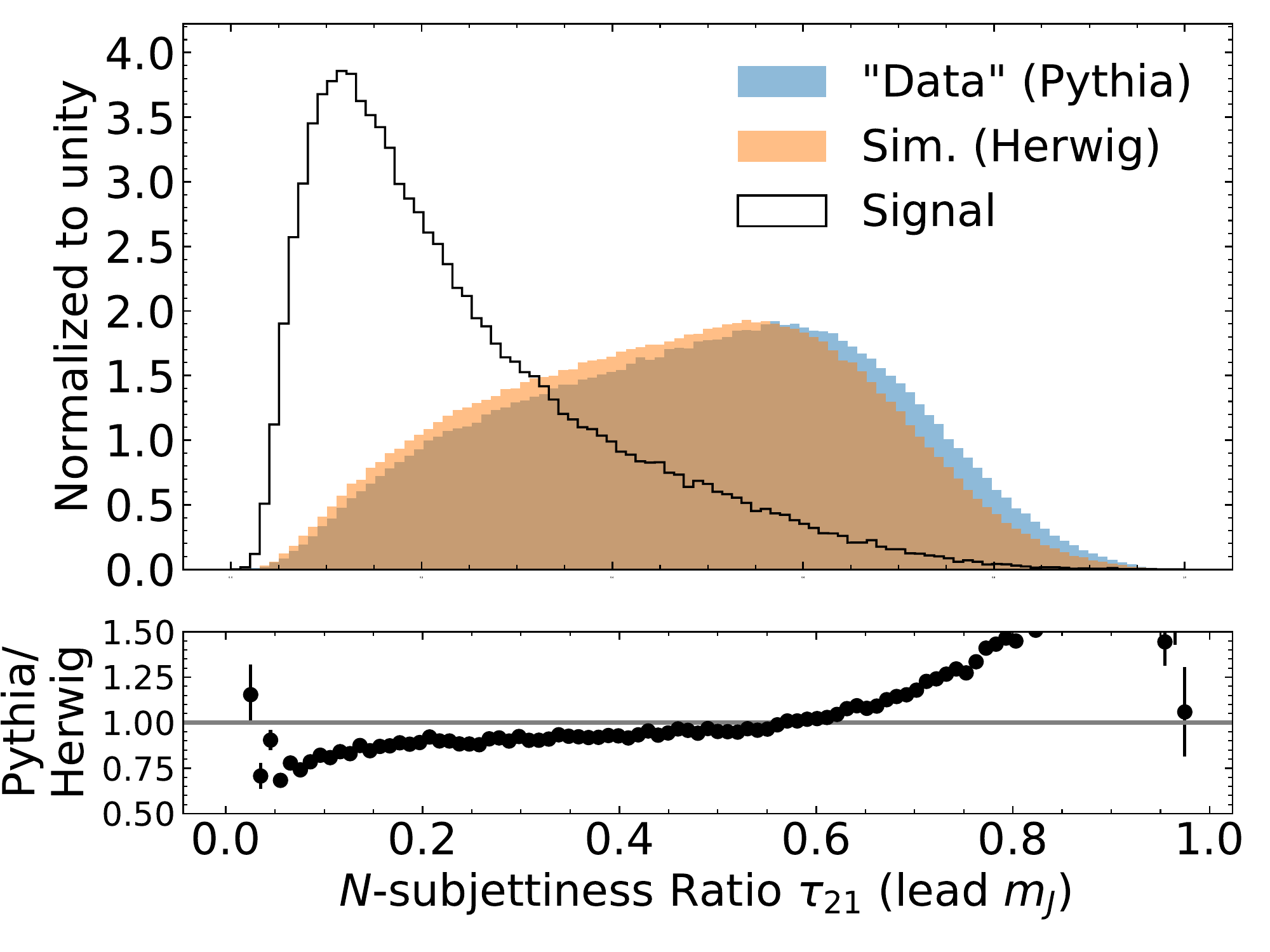}\includegraphics[width=0.5\textwidth]{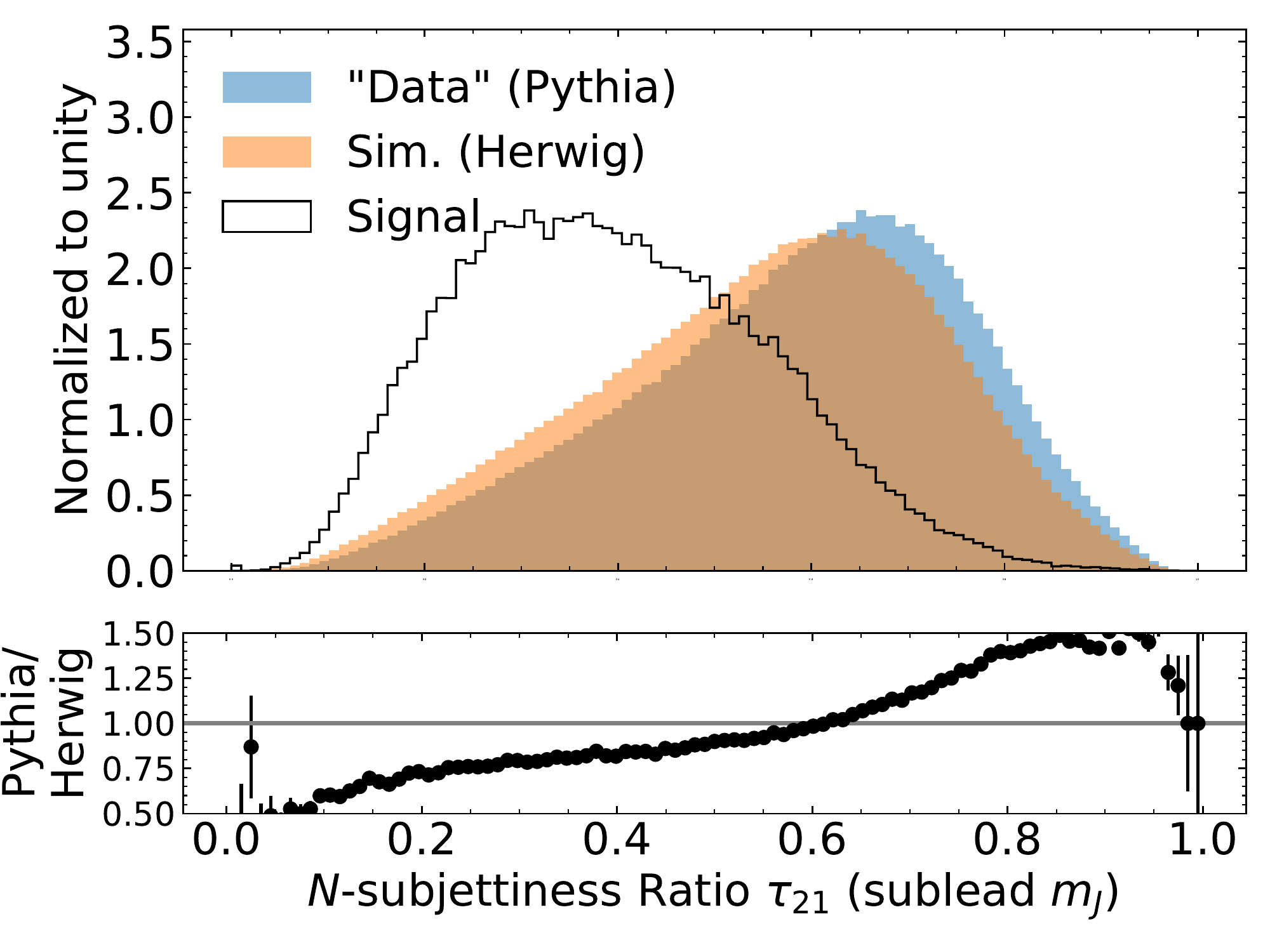}
\caption{The four features used for machine learning: jet mass (top) and the $N$-subjettiness ratios $\tau_{21}$ (bottom) for the more massive jet (left) and the less massive jet (right).}
\label{fig:mjetc}
\end{figure}

To demonstrate the \cathode~approach, two features\footnote{In principle, \cathode~can readily accommodate the full phase space as used by the original \dctr~method~\cite{Andreassen:2019nnm} based on particle flow networks~\cite{Komiske:2018cqr}; this will be explored in future studies.} about each of the leading jets are used for classification.  The first feature is the jet mass and the second feature is the $N$-subjettiness ratio~\cite{Thaler:2011gf,Thaler:2010tr} $\tau_{21}=\tau_2/\tau_1$.  This second feature is the most widely used feature for differentiating jets that have two hard prongs (as in the signal) from jets that have only one hard prong (as for most of the background).  The two jets are ordered by their mass and the four features used for machine learning are presented in Fig.~\ref{fig:mjetc}.  As expected, the signal mass distributions show peaks at the $X$ and $Y$ masses and the $\tau_{21}$ distributions are small, indicating two-prong substructure.  Pythia and Herwig differ mostly at low mass and across the entire $\tau_{21}$ distribution. 

\begin{figure}[h!]
\centering
\includegraphics[width=0.65\textwidth]{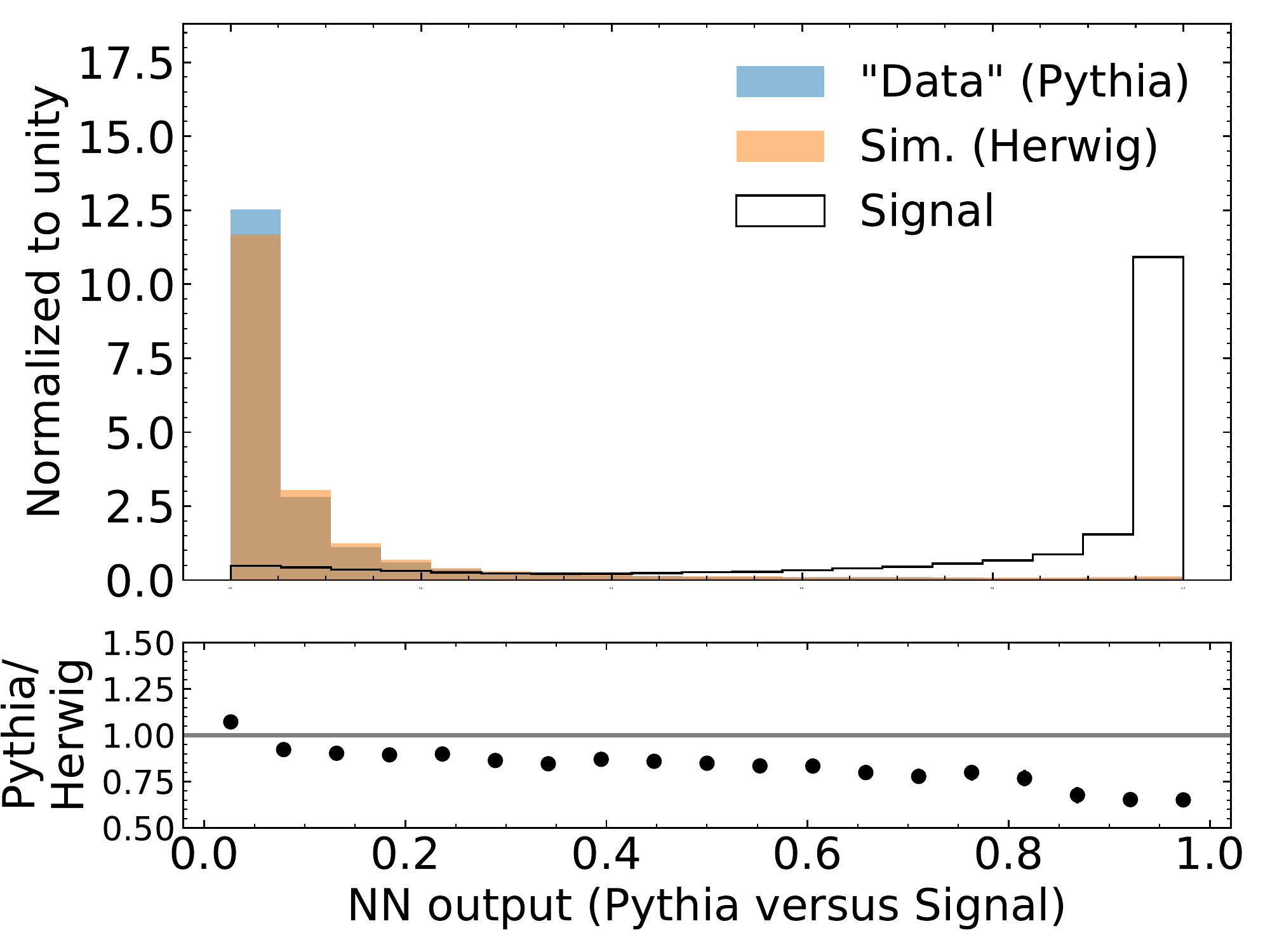}\\
\includegraphics[width=0.5\textwidth]{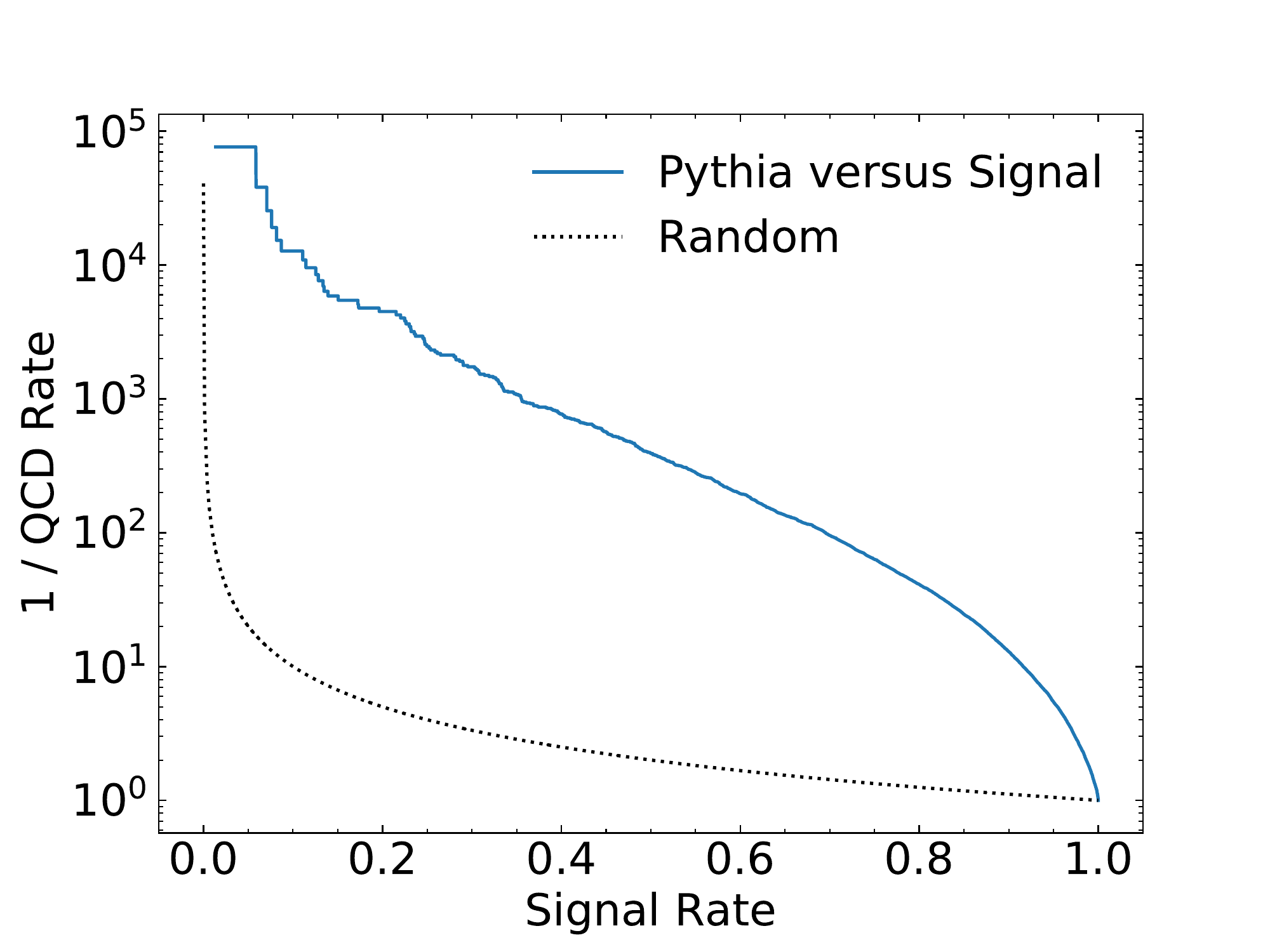}\includegraphics[width=0.5\textwidth]{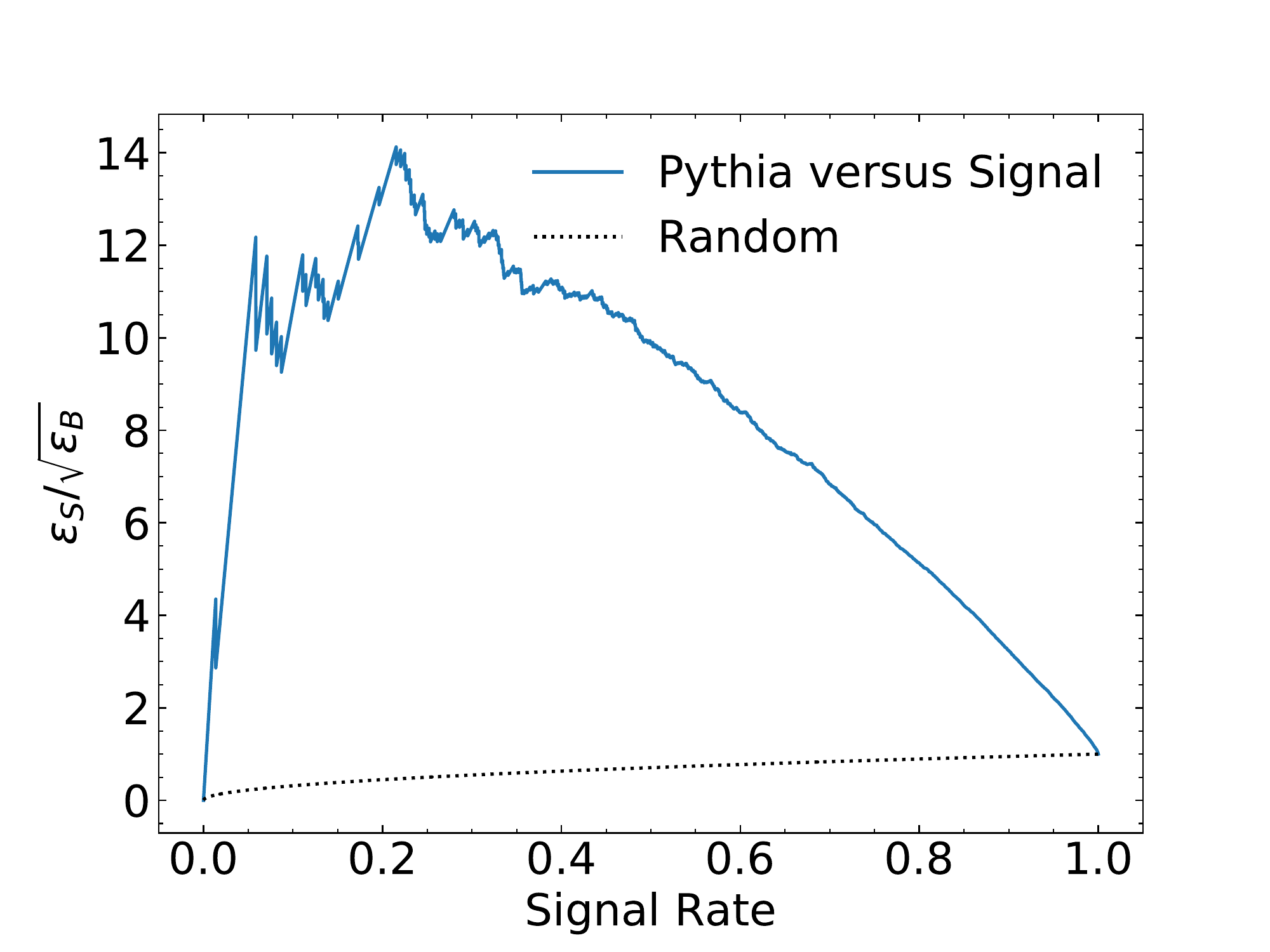}
\caption{A supervised classifier trained to distinguish signal from Pythia QCD.  The top plot is a histogram of the neural network output, the left bottom plot is a Receiver Operating Characteristic (ROC) curve, and the right bottom plot is a significance improvement (SIC) curve.  $\epsilon_S$ is the signal efficiency or true positive rate and $\epsilon_B$ is the background efficiency or false positive rate.}
\label{fig:superisedclassifier}
\end{figure}

The baseline performance for classifying signal versus the QCD background is presented in Fig.~\ref{fig:superisedclassifier}.    As is the case for all neural networks presented in the following sections, three fully connected layers with 100 hidden nodes on each intermediate layer are implemented using Keras~\cite{keras} and TensorFlow~\cite{tensorflow} with the Adam~\cite{adam} optimization algorithm.  Rectified linear units are the activation function for all intermediate layers and the sigmoid is used for the final output layer. Networks are trained with binary cross entropy for 50 epochs with early stopping (with patience 10).  The supervised classifier presented in Fig.~\ref{fig:superisedclassifier} effectively differentiates signal from background, with a maximum significance improvement of about 10.  It is expected that the performance of any model independent approach will be bounded from above by the performance of this classifier. 

\clearpage

\section{Parameterized Reweighting with DCTR}
\label{sec:dctr}

\begin{figure}[h!]
\centering
\includegraphics[width=0.65\textwidth]{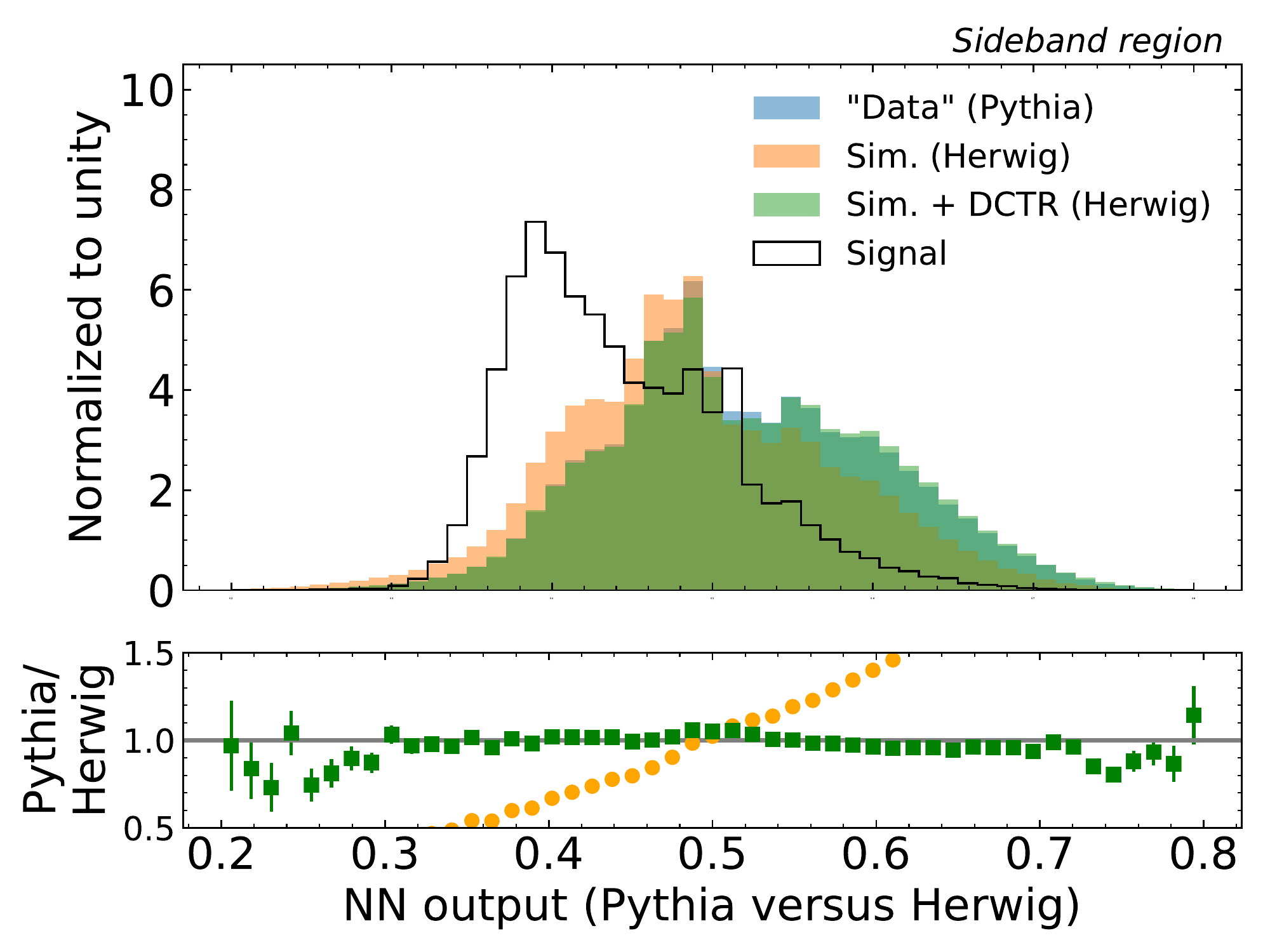}
\caption{A histogram of the classifier output for a neural network trained to distinguish `data' (Pythia) and `simulation' (Herwig) in the sideband region.}
\label{fig:dctrmodel}
\end{figure}

\begin{figure}[h!]
\centering
\includegraphics[width=0.5\textwidth]{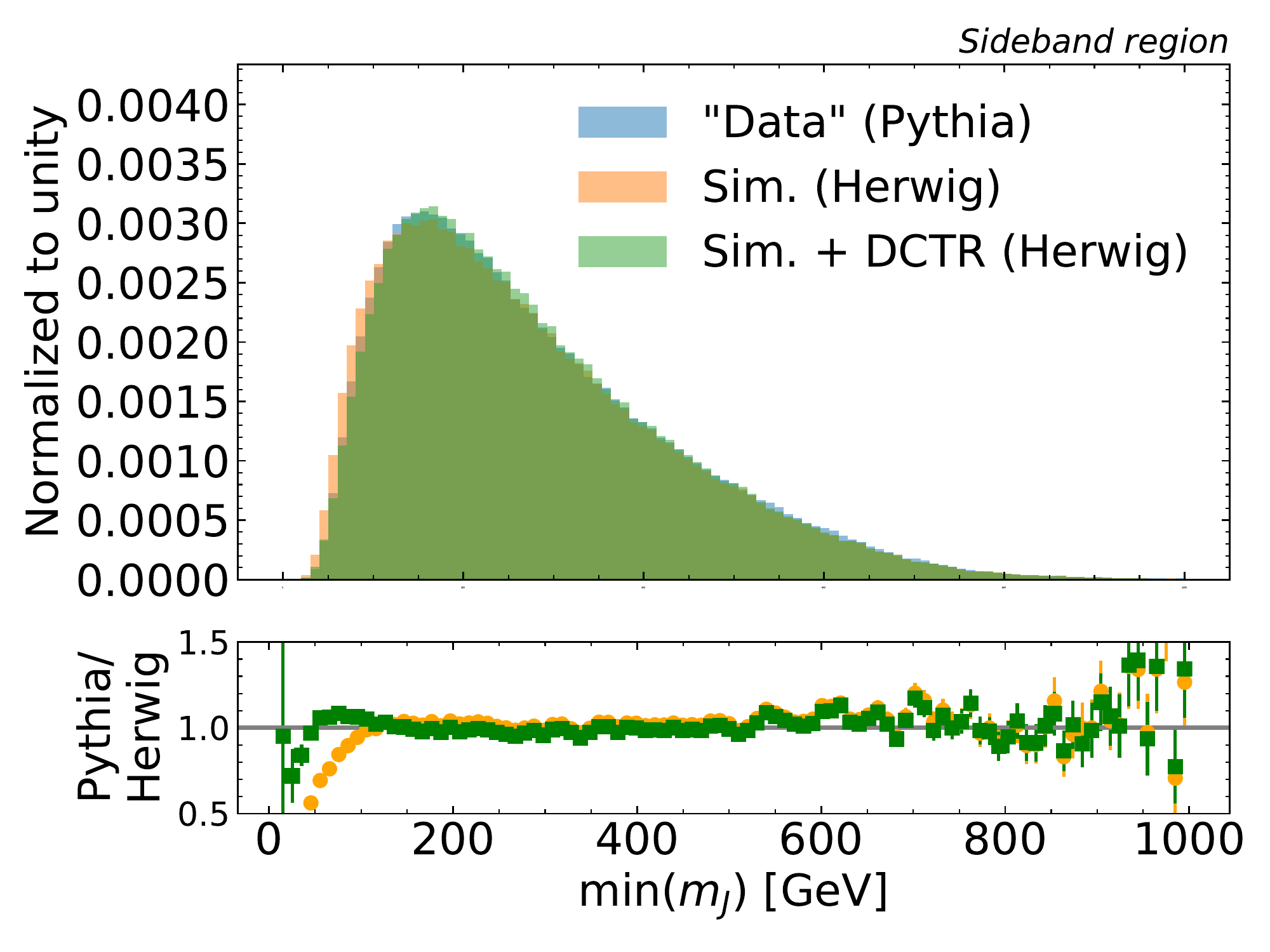}\includegraphics[width=0.5\textwidth]{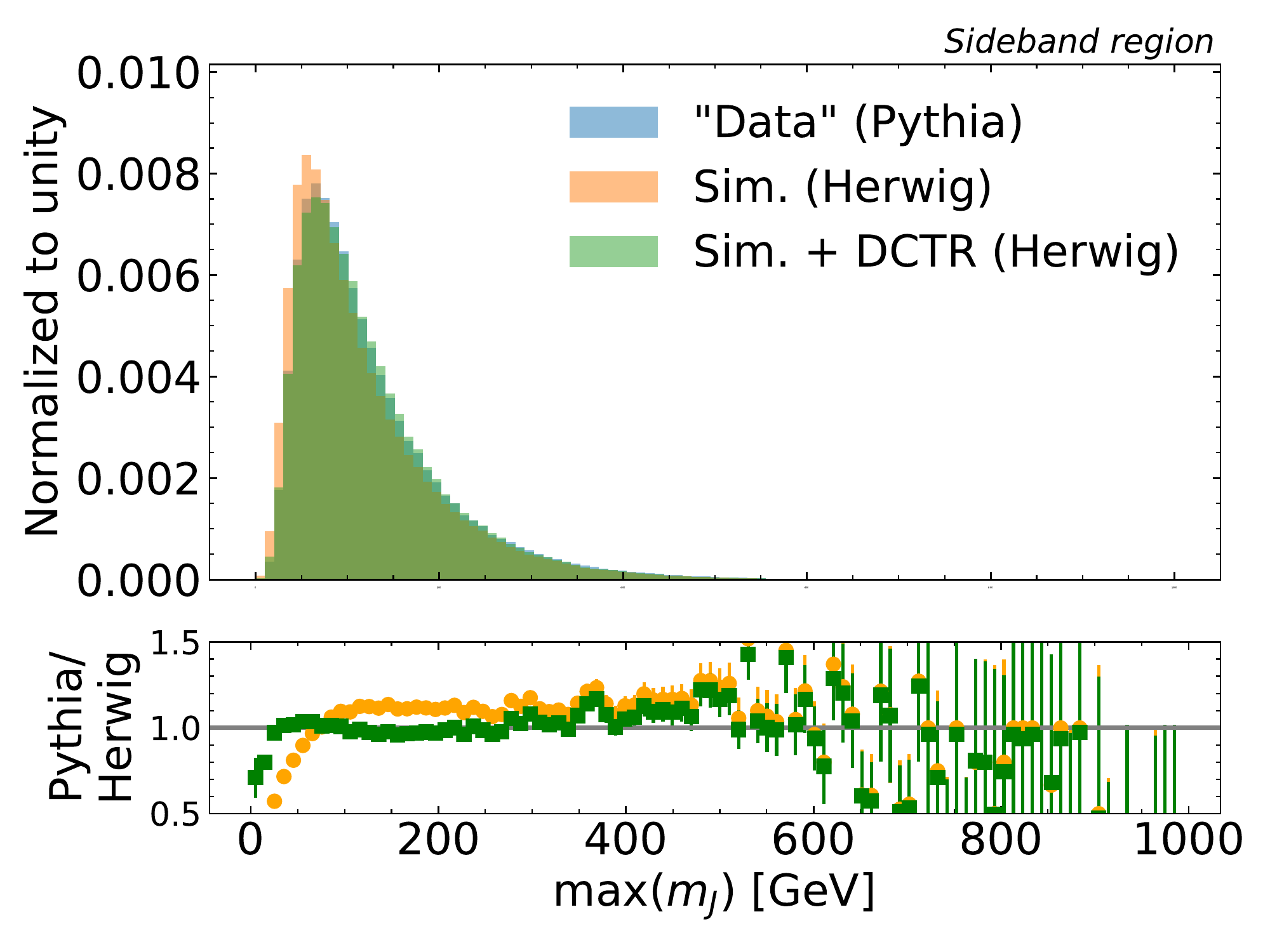}\\
\includegraphics[width=0.5\textwidth]{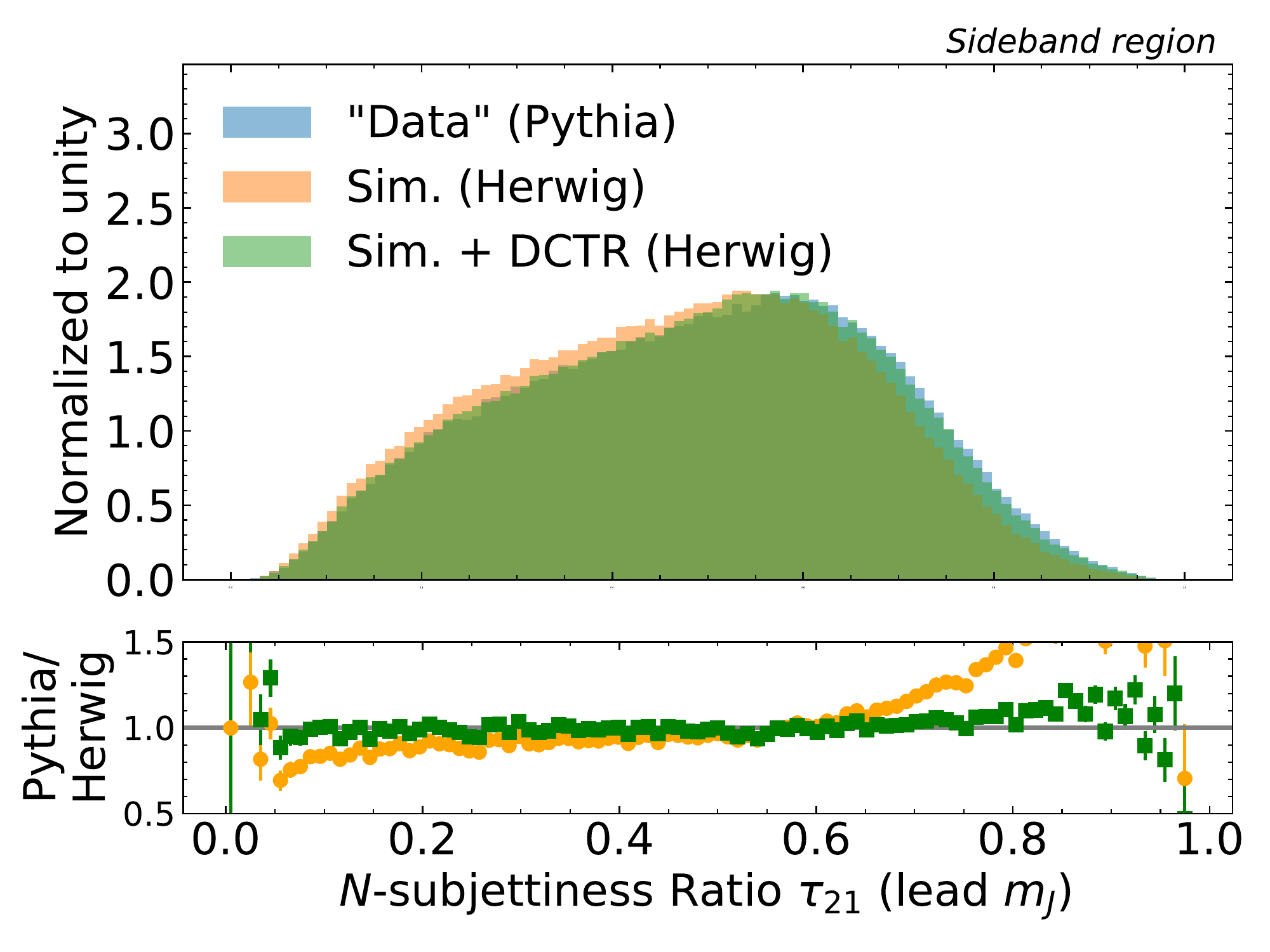}\includegraphics[width=0.5\textwidth]{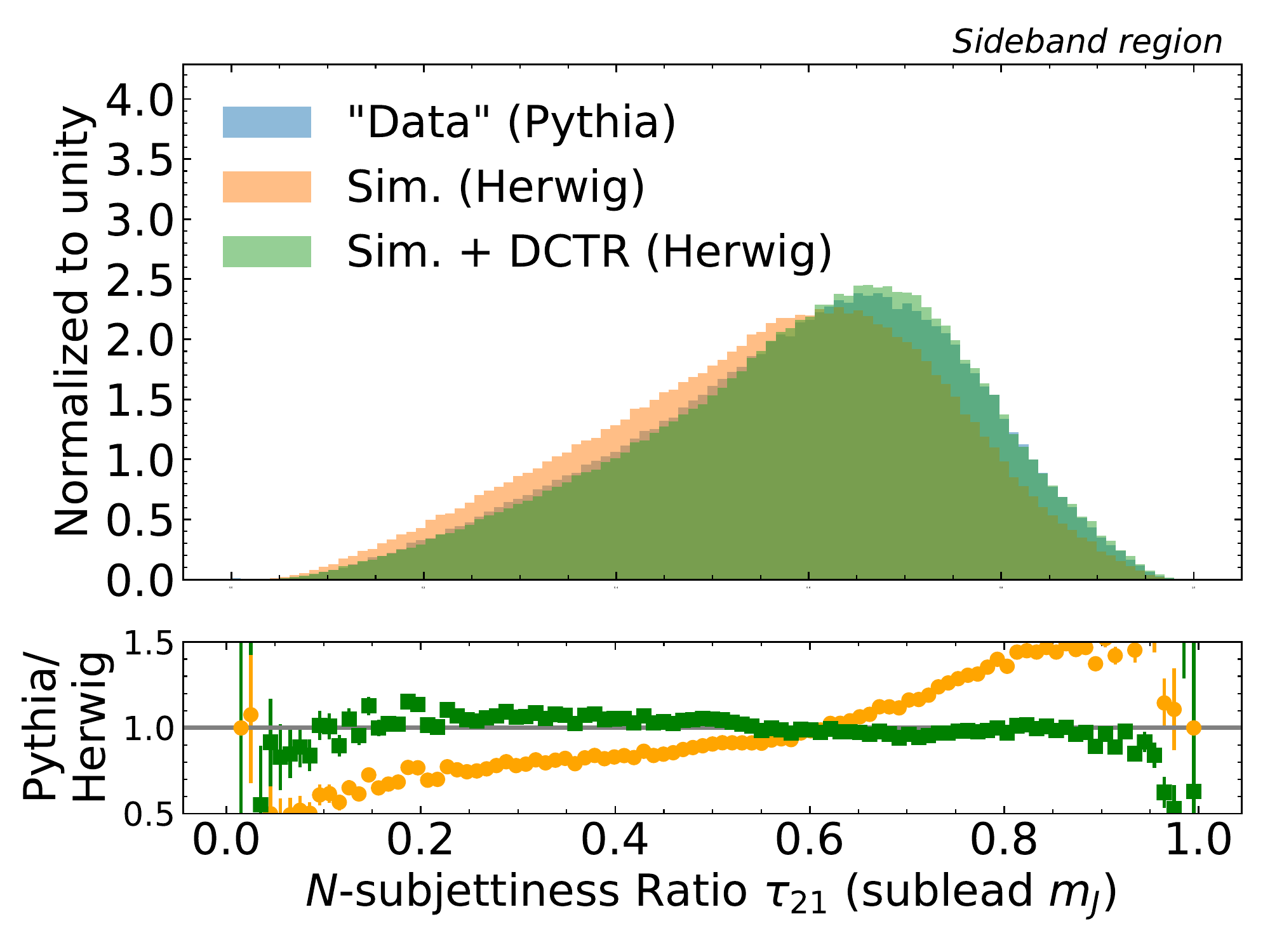}
\caption{The four features used for machine learning in the sideband region, before and after applying \dctr: jet mass (top) and the $N$-subjettiness ratios $\tau_{21}$ (bottom) for the more massive jet (left) and the less massive jet (right).}
\label{fig:mjetc_dctr_sb}
\end{figure}

The first step of the \dctr~reweighting procedure is to train a classifier to distinguish the `data' (Pythia) from the `simulation' (Herwig) in a sideband region.  The output of such a classifier is shown in Fig.~\ref{fig:dctrmodel}, where the signal region is defined as $m_{jj}\not\in[3250,3750]$ GeV.  There are about 850k events in the sideband region and 150k events in the signal region.  Unlike the classifier in Fig.~\ref{fig:superisedclassifier}, the separation in Fig.~\ref{fig:dctrmodel} is not as dramatic because Pythia and Herwig are much more similar than signal is with QCD.  As expected, the network is a linear function of the likelihood ratio so the ratio plot in Fig.~\ref{fig:dctrmodel} is linear.  Interestingly, the signal is more Herwig-like than Pythia-like.  The reweighting function is applied to the Herwig in Fig.~\ref{fig:dctrmodel} to show that the reweighted simulation (Sim.+\dctr) looks nearly identical to the `Data'.   All of the events used for Fig.~\ref{fig:dctrmodel} are independent from the ones used for training the network.  Figure~\ref{fig:mjetc_dctr_sb} shows shows that this reweighting works for all of the input distributions to the neural network as well.

\begin{figure}[h!]
\centering
\includegraphics[width=0.5\textwidth]{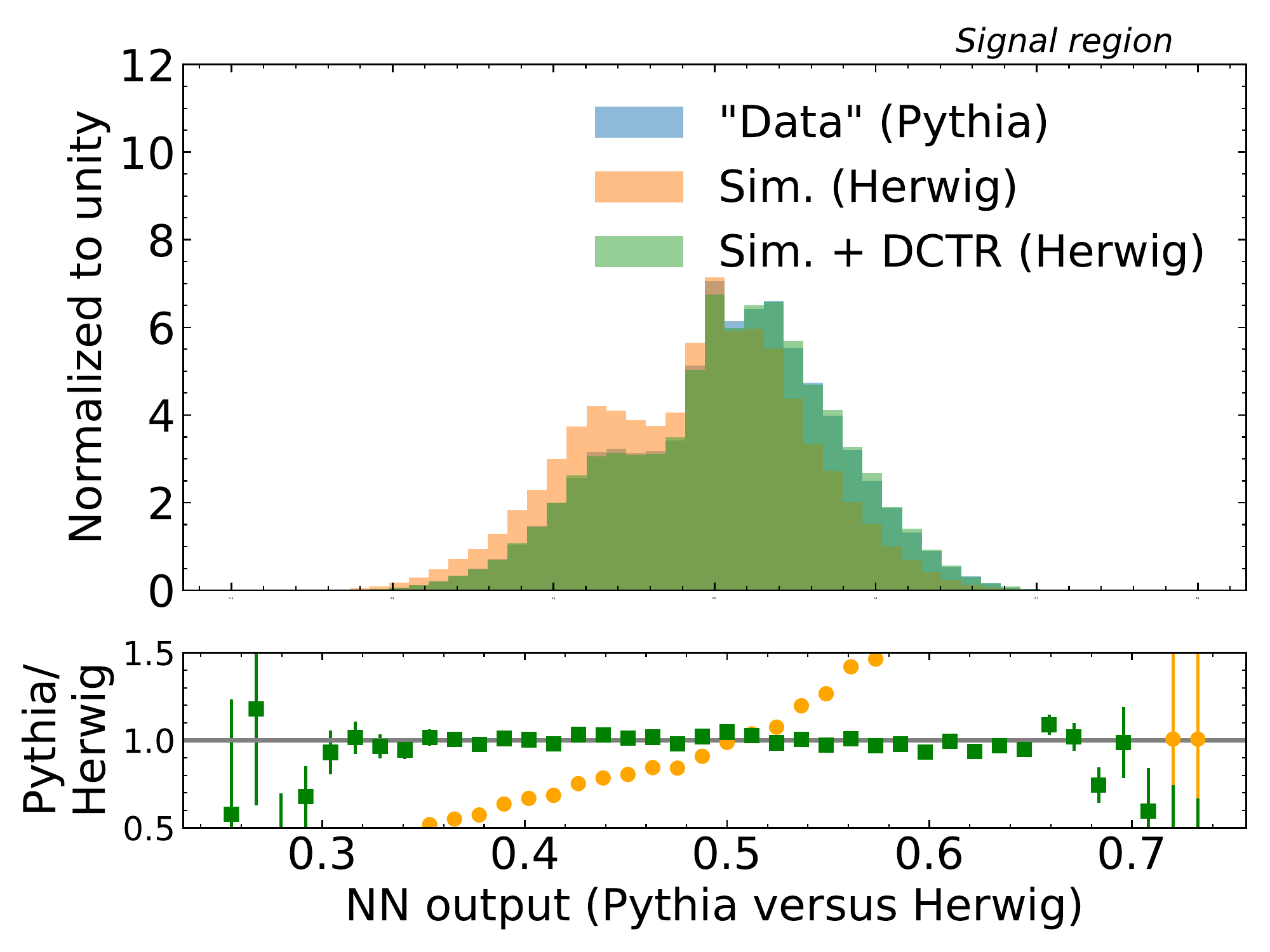}
\caption{A histogram of the classifier output for a neural network trained to distinguish `data' (Pythia) and `simulation' (Herwig) in the signal region.}
\label{fig:dctrmodel_SR}
\end{figure}

\begin{figure}[h!]
\centering
\includegraphics[width=0.45\textwidth]{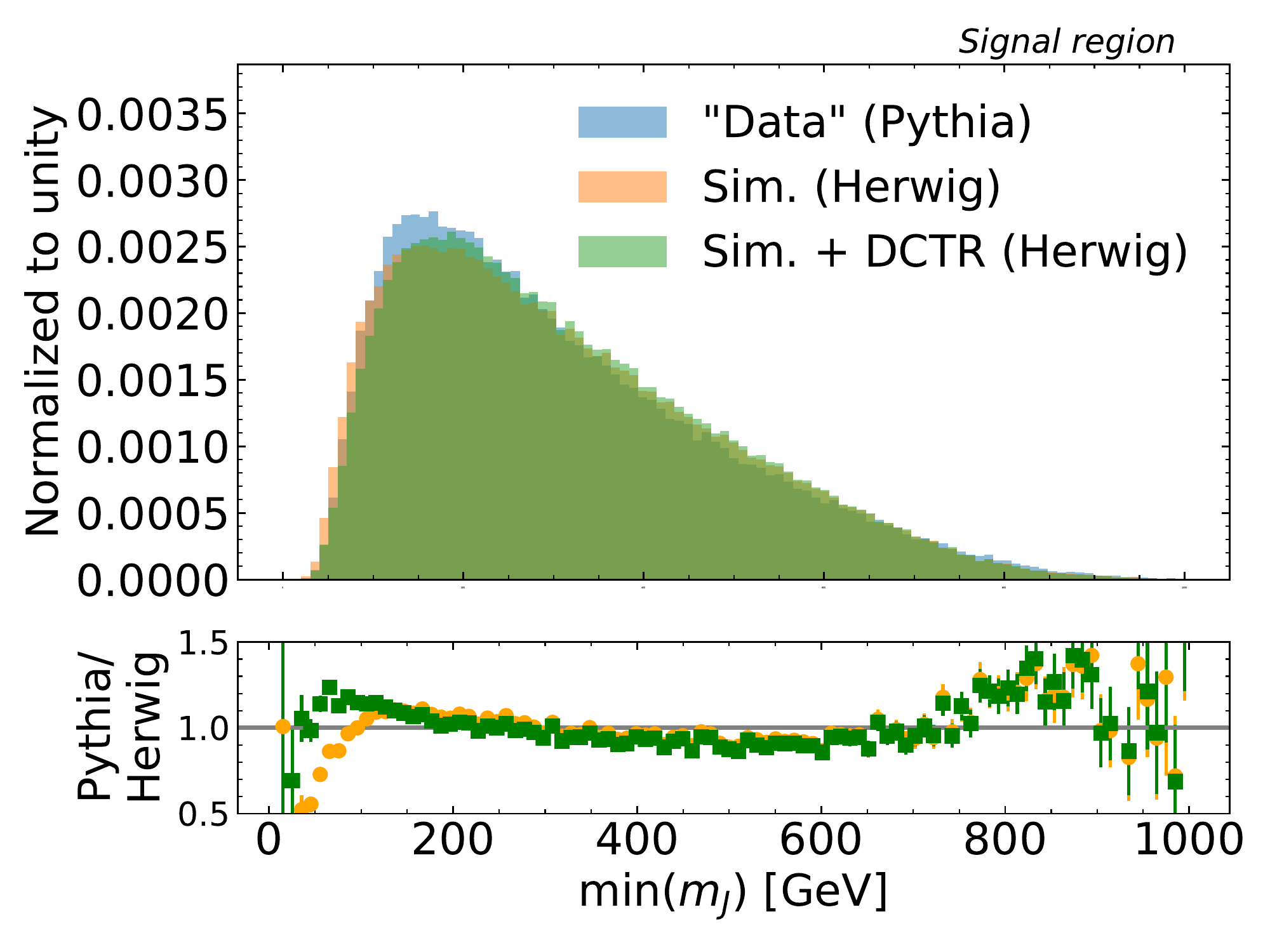}\includegraphics[width=0.45\textwidth]{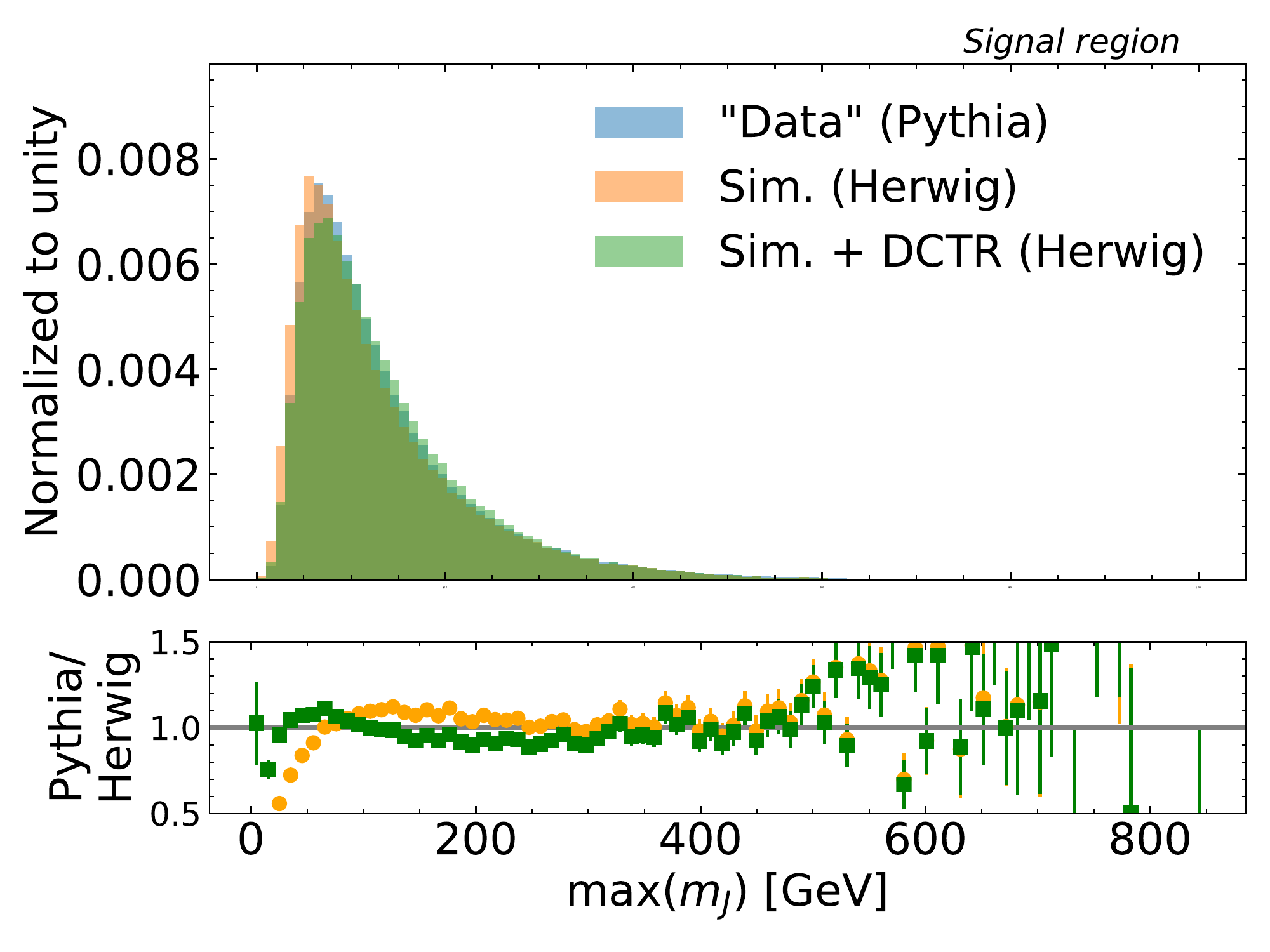}\\
\includegraphics[width=0.45\textwidth]{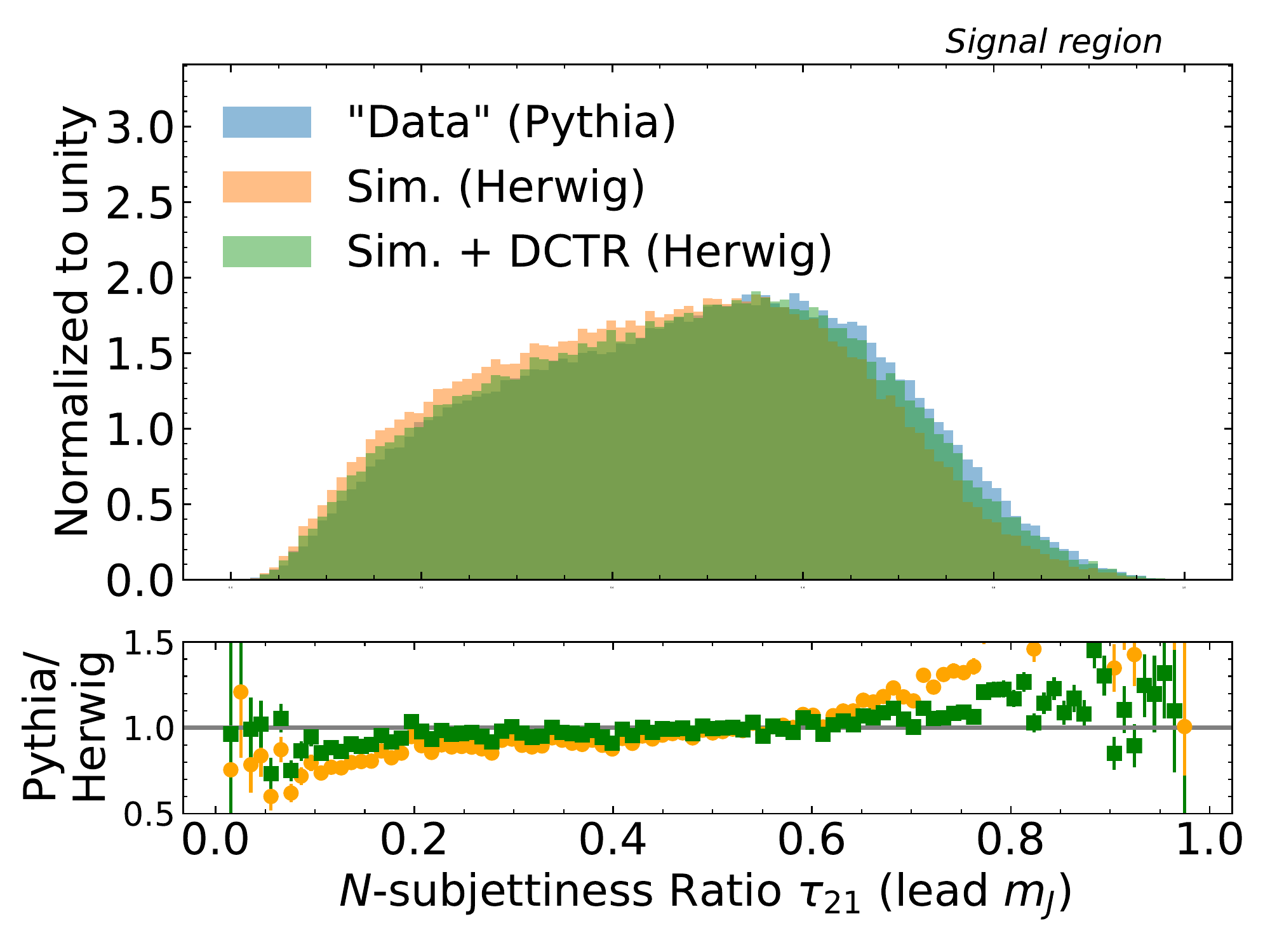}\includegraphics[width=0.45\textwidth]{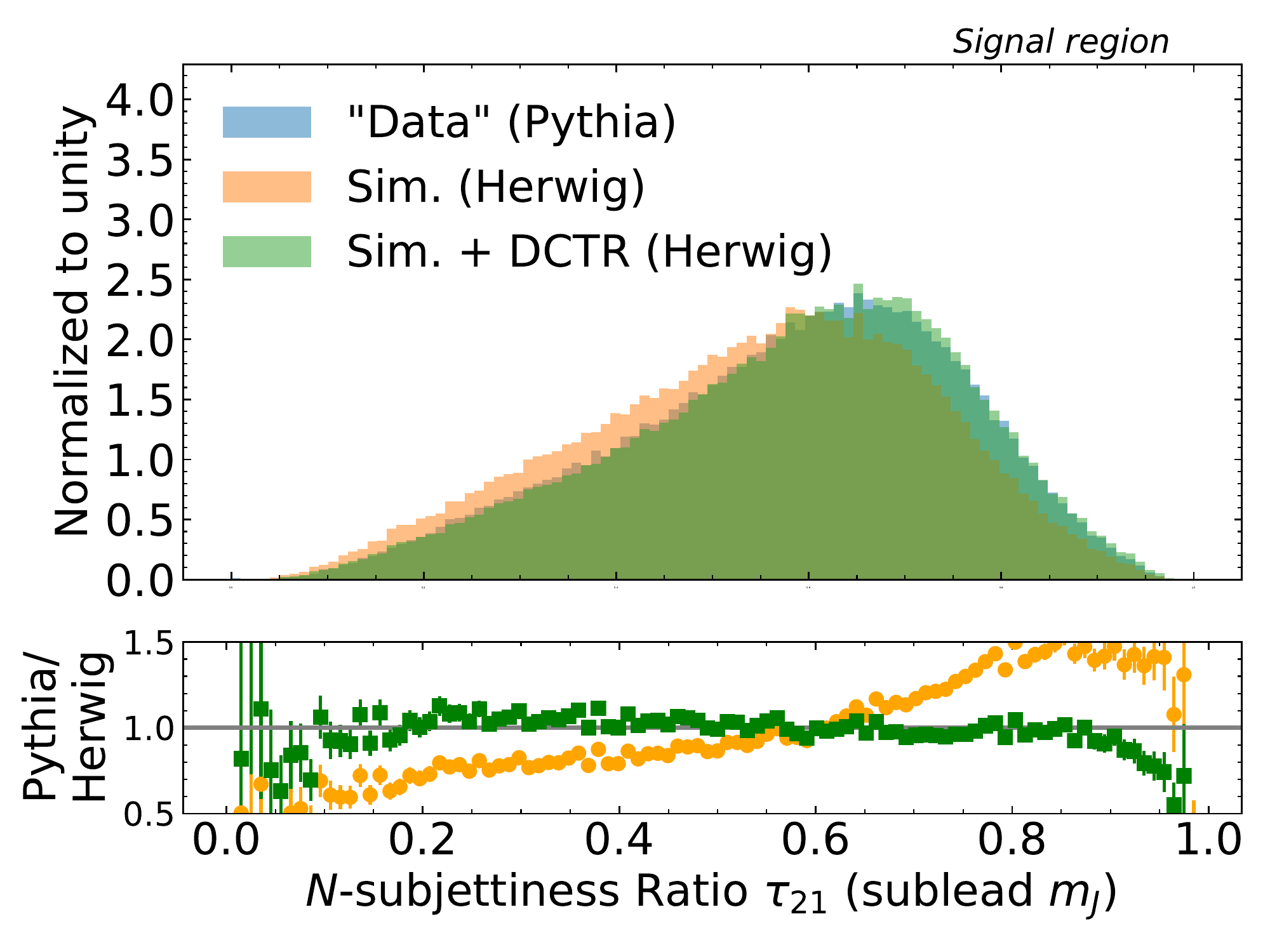}
\caption{The four features used for machine learning in the signal region, before and after applying \dctr: jet mass (top) and the $N$-subjettiness ratios $\tau_{21}$ (bottom) for the more massive jet (left) and the less massive jet (right).}
\label{fig:mjetc_dctr_sb_SR}
\end{figure}

The next step for \cathode~is to interpolate the reweighting function.  The neural network presented in Fig.~\ref{fig:dctrmodel} is trained conditional on $m_{jj}$ and so it can be evaluated in the SR for values of the invariant mass that were not available during the network training.  Note that the signal region must be chosen large enough so that the signal contamination in the sideband does not bias the reweighting function.  For this example, for 25\% signal fraction in the signal region, the contribution in the sideband is about 1\% and has no impact on the \dctr~model.  Figure~\ref{fig:dctrmodel_SR} shows a classifier trained to distinguish `data' and 'simulation' in the signal region before and after the application of the interpolated \dctr~model.  There is excellent closure, also for each of the input features to the classifier as shown in Fig.~\ref{fig:mjetc_dctr_sb_SR}.

\clearpage

\section{Sensitivity}
\label{sec:sens}

After reweighting the signal region to match the data, the next step of the search is to train a classifier to distinguish the reweighted simulation from the data in the signal region.  If the reweighting works exactly, then this new classifier will asymptotically learn $p(\text{signal}+\text{background})/p(\text{background})$, which is the optimal classifier by the Neyman-Pearson lemma~\cite{neyman1933ix}.  If the reweighting is suboptimal, then some of the classifier capacity will be diverted to learning the residual difference between the simulation and background data.  If the reweighted simulation is nothing like the data, then all of the capacity will go towards this task and it will not be able to identify the signal.  There is therefore a tradeoff between how different the (reweighted) simulation is from the data and how different the signal is from the background.  If the signal is much more different from the background than the simulation is from the background data, it is possible that a sub-optimally reweighted simulation will still be able to identify the signal (see Sec.~\ref{sec:back} for problems with background estimation).

\begin{figure}[h!]
\centering
\includegraphics[width=0.5\textwidth]{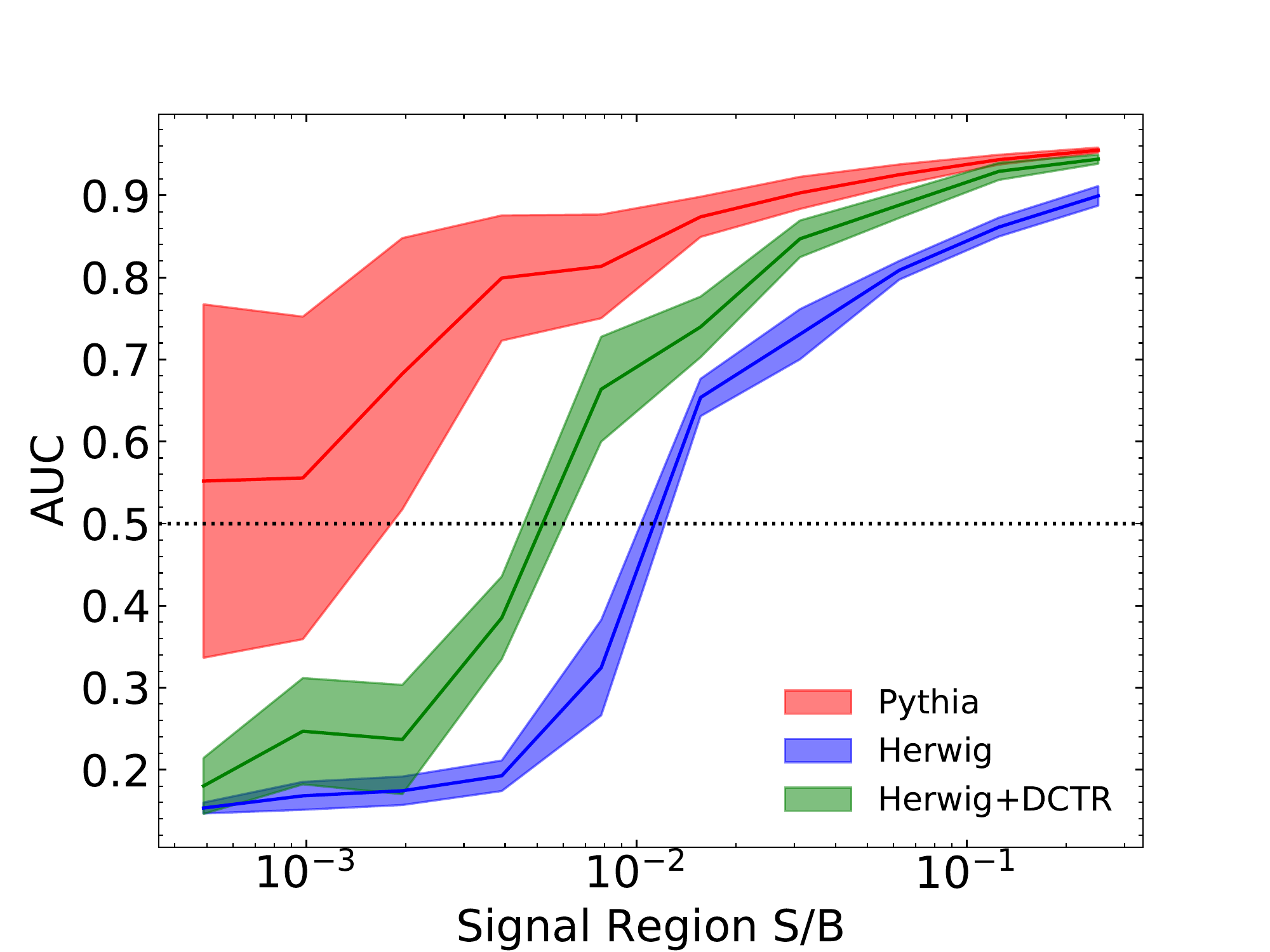}\includegraphics[width=0.5\textwidth]{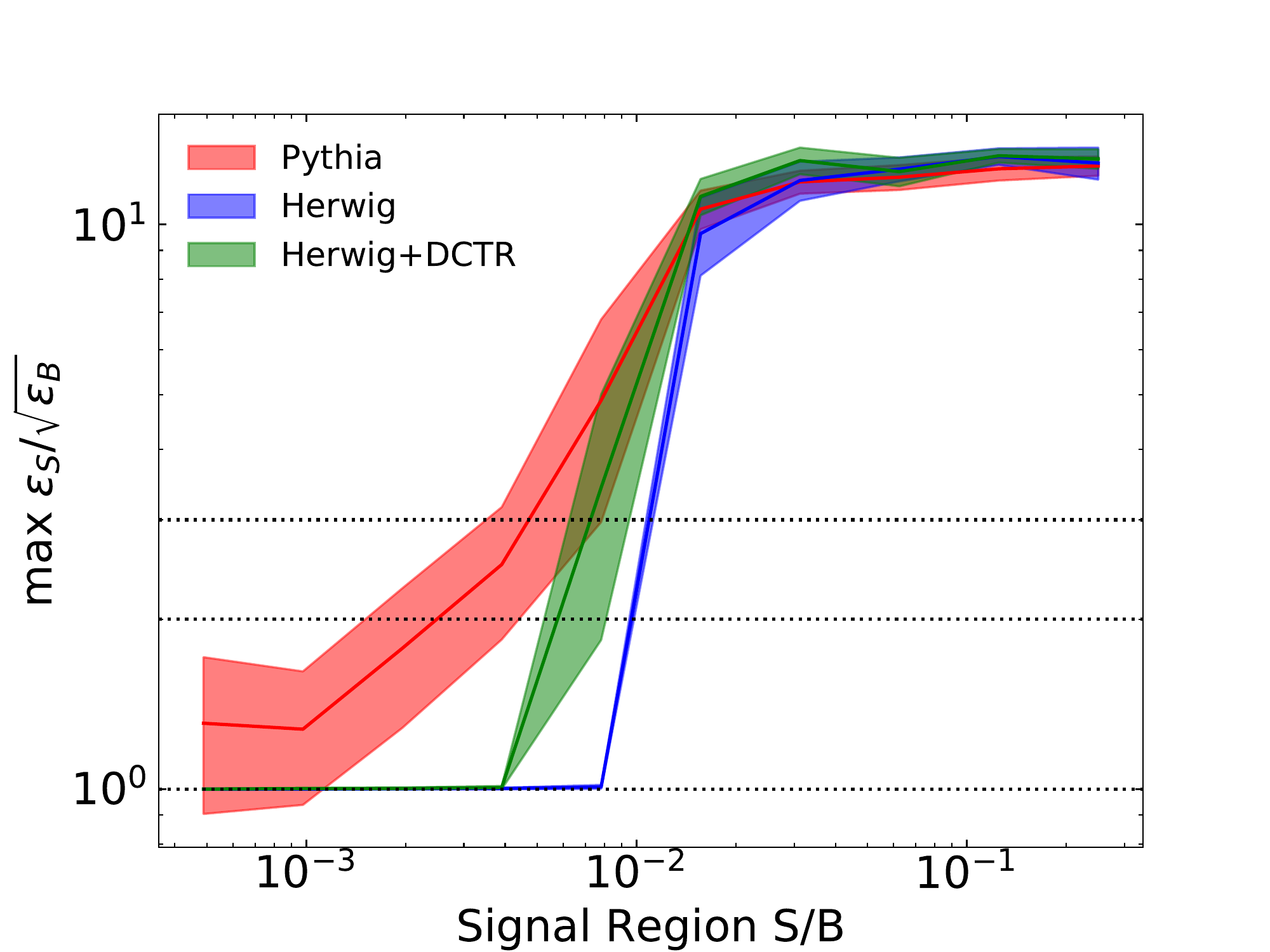}\\
\includegraphics[width=0.5\textwidth]{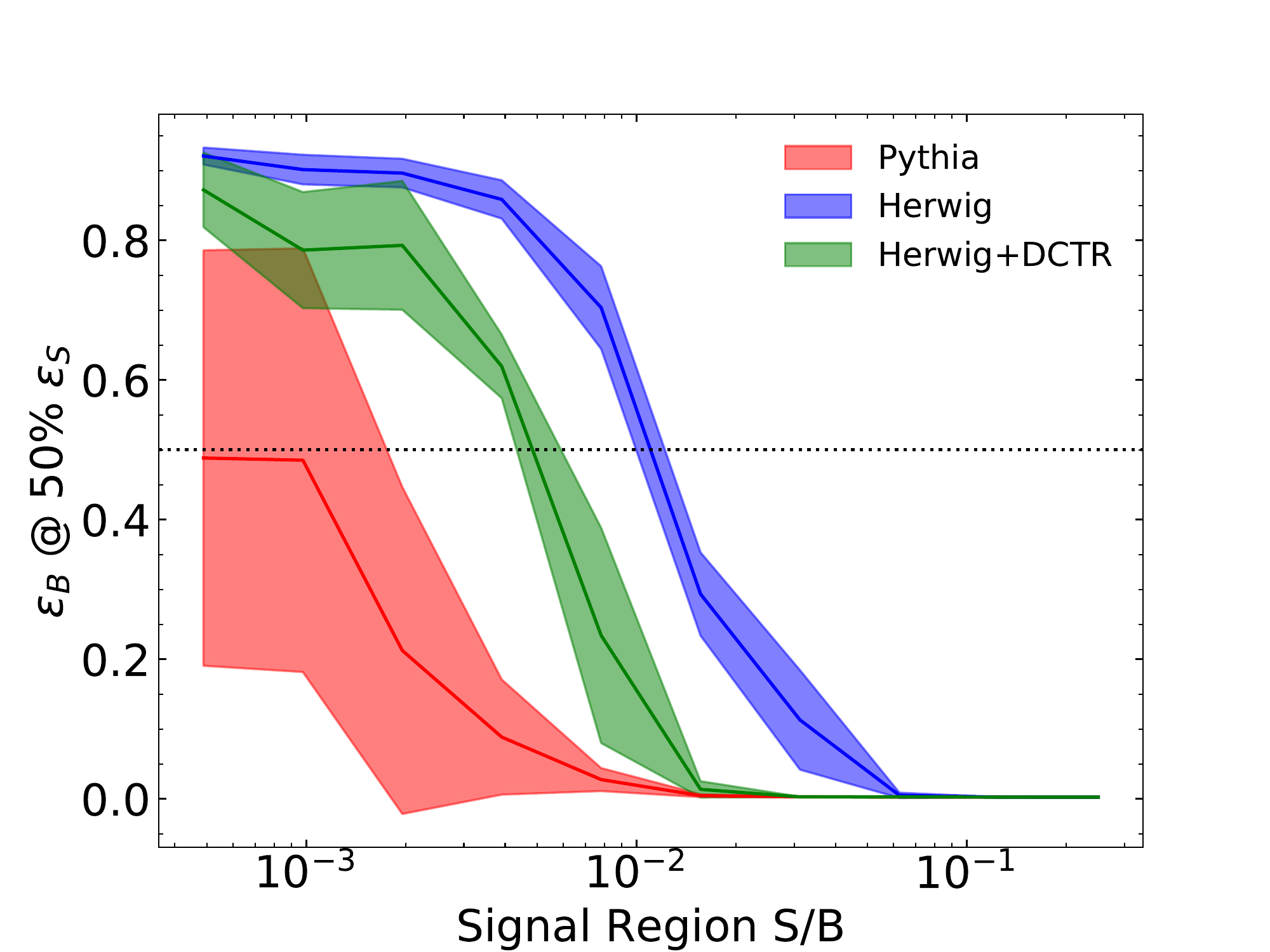}\includegraphics[width=0.5\textwidth]{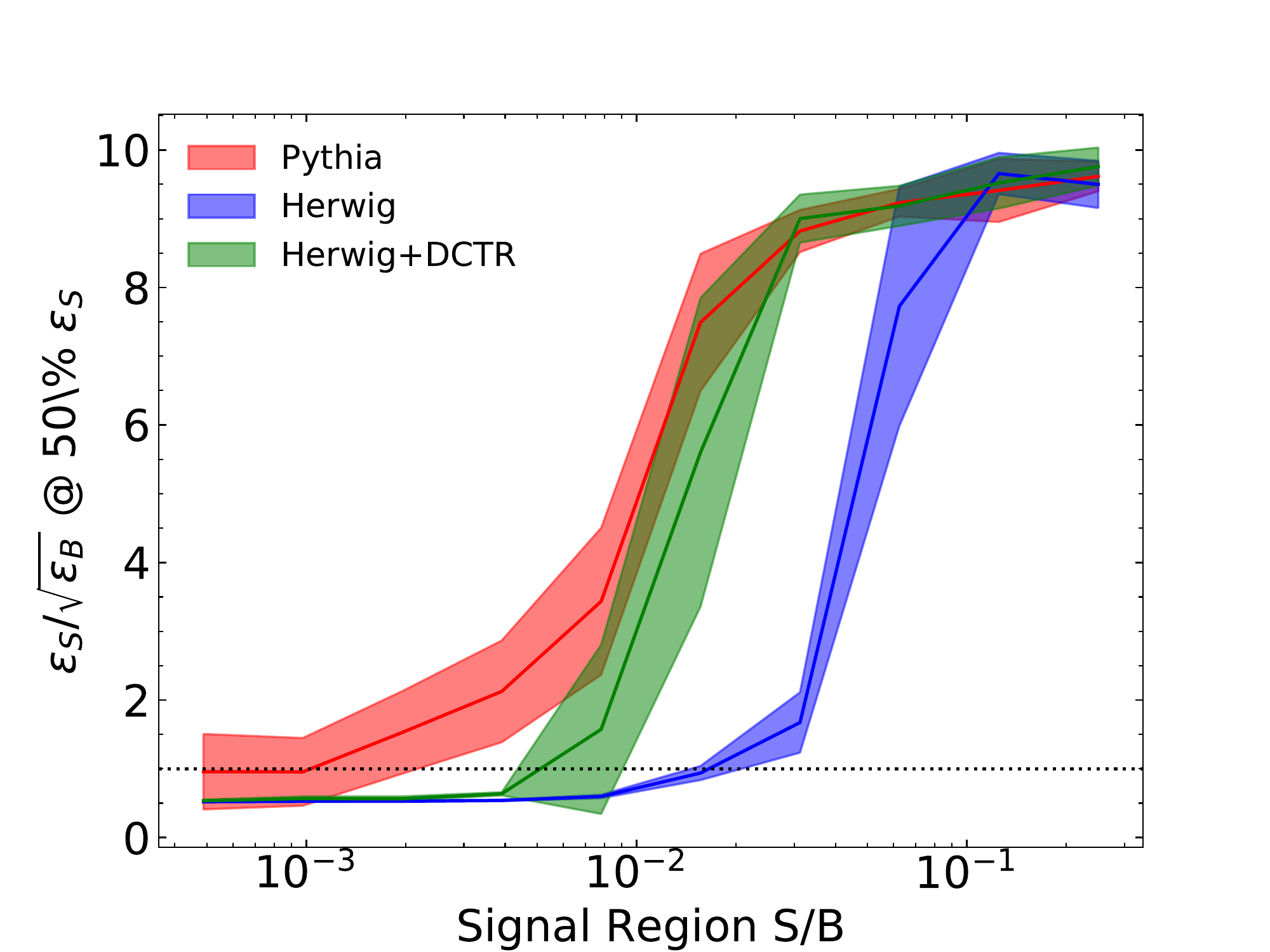}
\caption{Four metrics for the sensitivity of the \cathode~classifier as a function of the  signal-to-background ratio ($S/B$) in the signal region: the area under the curve (AUC) in the top left, the maximum significance improvement (top right), the false positive rate at a fixed 50\% signal efficiency (bottom left), and the significance improvement at the same fixed 50\% signal efficiency (bottom right).  The evaluation of these metrics requires signal labels, even though the training of the classifiers themselves do not have signal labels.  Error bars correspond to the standard deviation from training five different classifiers.  Each classifier is itself the truncated mean over ten random initializations.}
\label{fig:sensitivity}
\end{figure}

Figure~\ref{fig:sensitivity} shows the sensitivity of the \cathode~tagger to signal as a function of the signal-to-background ratio ($S/B$) in the signal region.  In all cases, the background is the QCD simulation using Pythia\footnote{Note that the full one million Pythia events are divided in two pieces, one that acts as the test set for all methods and one that is used for further study.  The remaining half is further split in half to represent the data or the simulation for the lines marked `Pythia' in Fig.~\ref{fig:sensitivity}.  For a fair comparison, the Herwig statistics are comparable to 25\% of the full Pythia dataset.  }.  The Pythia lines correspond to the case where the simulation follows the same statistics as the data ($=$ Pythia).   The area under the curve (AUC) should be as close to one as possible and a tagger that is operating uniformly at random will produce an AUC of 0.5.  Anti-tagging (preferentially tagging events that are not signal-like) results in an AUC less then 0.5.  The maximum significance improvement is calculated as the largest value of $\epsilon_S/\sqrt{\epsilon_B+0.01\%}$, where the 0.01\% offset regulates statistical fluctuations at low efficiency. 

When the $S/B\sim\mathcal{O}(1)$, then the performance in Fig.~\ref{fig:sensitivity} is similar to the fully supervised classifier presented in Sec.~\ref{sec:sim}.  As $S/B\rightarrow 0$, the Pythia curves approach the random classifier, with an AUC of 0.5 and a max significance improvement of unity.  The Herwig curve has an AUC less than 0.5 as $S/B\rightarrow 0$ because the signal is more Herwig-like than Pythia-like (see Fig.~\ref{fig:dctrmodel}) and thus a tagger that requires the features to be data-like (data $=$ Pythia) will anti-tag the signal.  Likewise, the efficiency of the tagger on the simulation is higher than 50\% when placing a threshold on the NN that keeps 50\% of the events in data.  The maximum significance improvement quickly drops to unity for Herwig when $S/B\lesssim 1\%$, indicating the the network is spending more capacity on differentiating Pythia from Herwig than finding signal.

For all four metrics, \cathode~significantly improves the performance of the Herwig-only approach.  In particular, the \cathode~tagger is effective to about $S/B\lesssim 0.5\%$, whereas the Herwig-only tagger is only able to provide useful discrimination power down to about $S/B\sim 1\%$.   For the significance improvement and false positive rate at a fixed true positive rate, the \cathode~tagger tracks the Pythia tagger almost exactly down to below 1\%.  The AUC about half way between Pythia and Herwig at high $S/B$, which is indicative of poor performance at low efficiency.  

\section{Background Estimation}
\label{sec:back}

\begin{figure}[h!]
\centering
\includegraphics[width=0.65\textwidth]{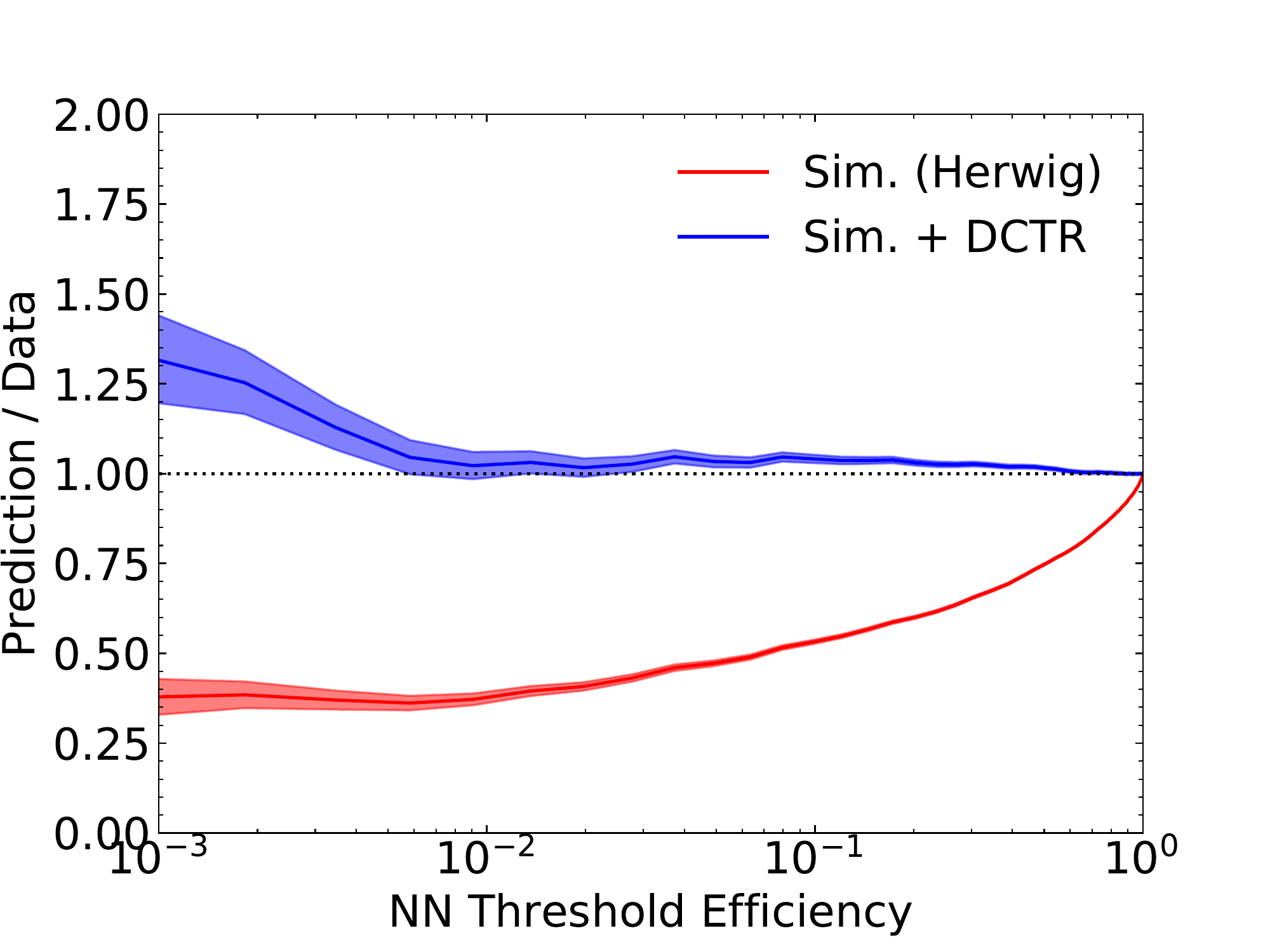}
\caption{The predicted efficiency normalized to the true data efficiency in the signal region for various threshold requirements on the NN.  The $x$-axis is the data efficiency from the threshold.  The error bars are due to statistical uncertainties.}
\label{fig:specificity}
\end{figure}

The performance gains from Sec.~\ref{sec:sens} can be combined with a sideband background estimation strategy, as long as threshold requirements on the classifier do not sculpt bumps in the $m_{jj}$ spectrum.  However, there is also an opportunity to use \cathode~to directly estimate the background from the interpolated simulation.  Figure~\ref{fig:specificity} illustrates the efficacy of the background estimation for a single classifier trained in the absence of signal.  Without the \dctr~reweighting, the predicted background rate is too low by a factor of two or more below 10\% data efficiency.  With the interpolated reweighting function, the background prediction is accurate within a few percent down to about 1\% data efficiency.

In practice, the difficulty in using \cathode~to directly estimate the background is the estimation of the residual bias.  One may be able to use validation regions between the signal region and sideband region, but it will never require as much interpolation as the signal region itself.  One can rely on simulation variations and auxiliary measurements to estimate the systematic uncertainty from the direct \cathode~background estimation, but estimating high-dimensional uncertainties is challenging~\cite{Nachman:2019dol,Nachman:2019yfl}.   With a low-dimensional reweighting or with a proper high-dimensional systematic uncertainty estimate, the parameterized reweighting used in \cathode~should result in a lower uncertainty than directly estimating the uncertainty from simulation.  In particular, any nuisance parameters that affect the sideband region and the signal region in the same way will cancel when reweighting and interpolating.

\clearpage

\section{Conclusions}
\label{sec:conclusions}

This paper has introduced \textit{Simulation Assisted Likelihood-free Anomaly Detection} (\cathode), a new approach to search for resonant anomalies by using parameterized reweighting functions for classification and background estimation.  The \cathode~approach uses information from simulation in a way that is nearly background-model independent while remaining signal-model agnostic.  The only requirement for the signal is that there is one feature where the signal is known to be localized.  In the example presented in the paper, this feature was the invariant mass of two jets.  The location of the resonance need not be known ahead of time and can be scanned using a series of signal and sideband regions.  This scanning will result in a trials factor per non-overlapping signal region.  An additional look elsewhere effect is incurred by scanning the threshold on the neural network.  In practice, one could use a small number of widely separated thresholds to be broadly sensitive.  As long as the data used for training and testing are independent, there is no additional trials factor for the feature space used for classification.  Strategies for maximally using the data for training can be found in Ref.~\cite{Collins:2018epr,Collins:2019jip}.

While the numerical \cathode~results presented here did not fully achieve the performance of a fully supervised classifier trained directly with inside knowledge about the data, there is room for improvement.  In particular, a detailed hyperparameter scan could improve the quality of the reweighting.  Additionally, calibration techniques could be used to further increase the accuracy~\cite{Cranmer:2015bka}.  Future work will investigate the potential of \cathode~to analyze higher-dimensional feature spaces as well as classifier features that are strongly correlated with the resonant feature.  It will also be interesting to compare \cathode~with other recently proposed model independent methods.  When the nominal background simulation is an excellent model of nature, \cathode~should perform similarly to the methods presented in Ref.~\cite{DAgnolo:2018cun,DAgnolo:2019vbw} and provide a strong sensitivity to new particles.  In other regimes where the background simulation is biased, \cathode~should continue to provide a physics-informed but still mostly background/signal model-independent approach to extend the search program for new particles at the LHC and beyond.

\section*{Code and data availability}

The code can be found at \url{https://github.com/bnachman/DCTRHunting} and the datasets are available on Zendo as part of the LHC Olympics~\cite{gregor_kasieczka_2019_2629073}.

\acknowledgments

BN would like to thank Jack Collins for countless discussions about anomaly detection, including ideas related to \cathode.   Some of those discussions happened at the Aspen Center for Physics, which is supported by National Science Foundation grant PHY-1607611.  This work was supported by the U.S.~Department of Energy, Office of Science under contract DE-AC02-05CH11231.   DS is supported by DOE grant DOE-SC0010008.  DS thanks LBNL, BCTP and BCCP for their generous support and hospitality during his sabbatical year.   BN would like to thank NVIDIA for providing Volta GPUs for neural network training.

\bibliographystyle{jhep}
\bibliography{myrefs}
\end{document}